\documentclass[twocolumn]{article}
 \usepackage{authblk}
 
\usepackage[right]{lineno}
\usepackage{amsmath,amsfonts}
\usepackage{algorithmic}
\usepackage{algorithm}
\usepackage{array}
\usepackage[caption=false,font=normalsize,labelfont=sf,textfont=sf]{subfig}
\usepackage{textcomp}
\usepackage{stfloats}
\usepackage{url}
\usepackage{verbatim}
\usepackage{graphicx}
\usepackage{hyphenat}
\usepackage{epstopdf}

\DeclareMathOperator{\sinc}{sinc}

\begin{document}

\title{{Real-time auralization for performers on virtual stages}}

\author[1]{Ernesto Accolti}
\author[2]{Lukas Aspöck}
\author[2]{Manuj Yadav}
\author[2]{Michael Vorländer}
\affil[1]{Instituto de Autom\'atica (Institute of Automation), National University of San Juan (UNSJ) and the National Scientific and Technical Research Council (CONICET), Argentina}
\affil[2]{Institute for Hearing Technology and Acoustics, RWTH Aachen University, Aachen, Germany}
   \maketitle

\begin{abstract}
This article presents an interactive system for stage acoustics experimentation including considerations for hearing one's own and others' instruments. 
The quality of real-time auralization systems for psychophysical experiments on music performance depends on the system's calibration and latency, among other factors (e.g. visuals, simulation methods, haptics, etc). The presented system focuses on the acoustic considerations for laboratory implementations. 
The calibration is implemented as a set of filters accounting for the microphone-instrument distances and the directivity factors, as well as the transducers' frequency responses. 
Moreover, sources of errors are characterized using both state-of-the-art information and derivations from the mathematical definition of the calibration filter. 
In order to compensate for hardware latency without cropping parts of the simulated impulse responses, the virtual direct sound of musicians hearing themselves is skipped from the simulation and addressed by letting the actual direct sound reach the listener through open headphones. 
The required latency compensation of the interactive part (i.e. hearing others) meets the minimum distance requirement between musicians, which is 2 m for the implemented system. 
Finally, a proof of concept is provided that includes objective and subjective experiments, which give support to the feasibility of the proposed setup.
\end{abstract}

\section{Introduction}
\label{sec:Introduction}

Auralization has a long tradition \cite{lehnert1992,kleiner1993,blauert1997}. 
Furthermore, it has greatly developed in the last decade
\cite{Kyoungsoo2014614, Funkhouser1998, Savioja1999,Shelley2010,ahrens2018, Llorca2021,vorlander_auralizatio_2020,Firat2022}. 
Although realism has reached a great state of development,  also for inclusion of musical instrument directivities \cite{Kyoungsoo2014614}, many improvements are still possible \cite{Blau2021}.

Currently, auralization with high plausibility  is emerging in many applications -- some even involving real-time processing -- primarily for cases where the sources are external to the listener, and\fshyp{}or relatively far away. However, some details still remain as open research topics. That is especially the case of real-time auralization for applications involving musicians hearing themselves and/or their own instruments, musicians hearing other musicians\fshyp{}instruments 
who are very close, or other applications involving short source\hyp{}receiver distances \cite{p-g_equal_2011,y_c_m_2012,gari2017,yadav2017,kim2020,kittimathaveenan2021}. A similar scheme but intended for voice communication was introduced by \cite{huang2011} as an overview of 
services infrastructure 
including multiparty immersive audio mixing and management, as well as immersive sound rendering.

\begin{figure}[b!]
	\centering
		\includegraphics[width=0.45\textwidth]{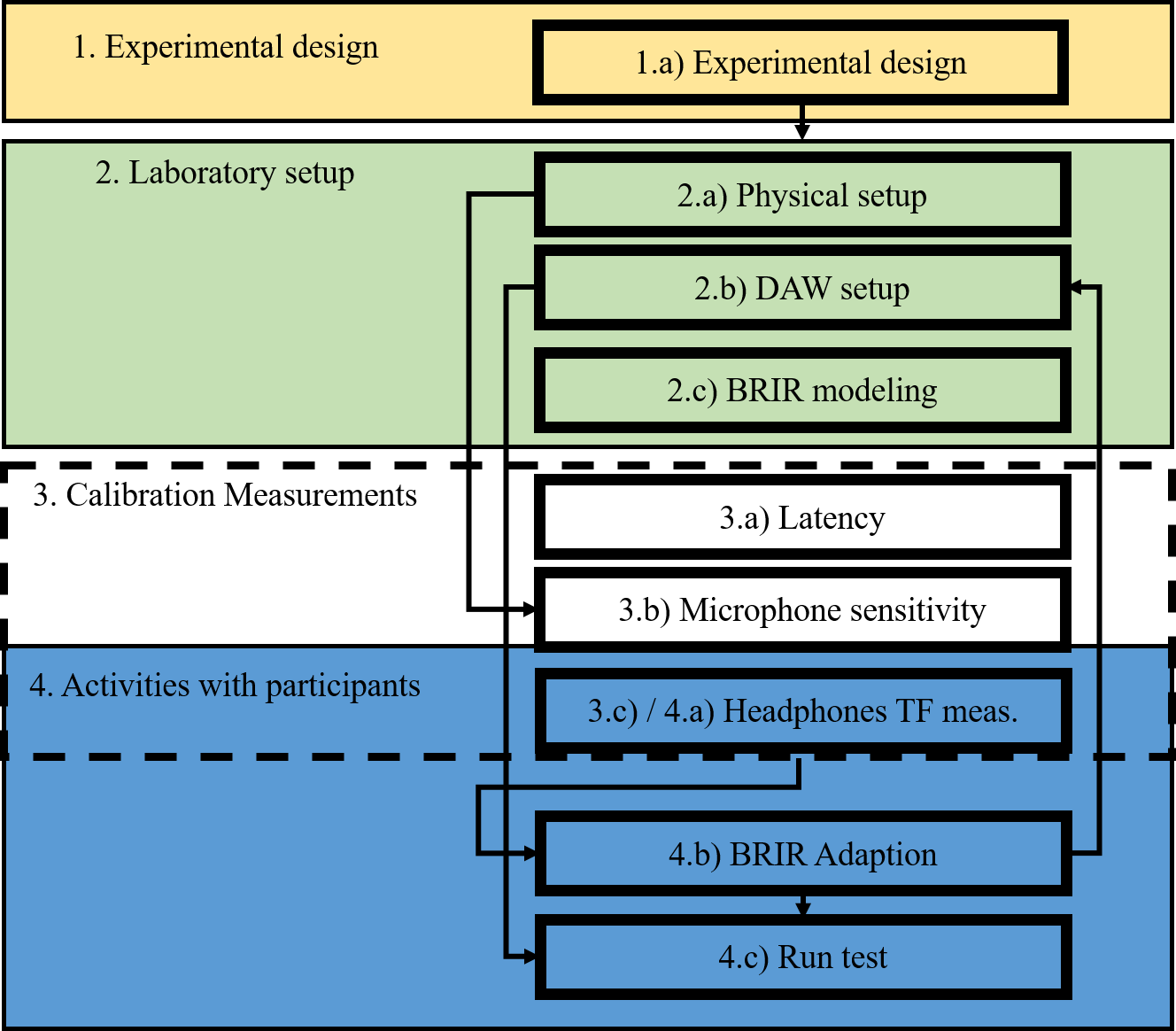}
\caption{\label{fig:auralization_scheme} Auralization scheme. DAW: Digital audio workstation, BRIR: Binaural room Impulse Response, TF meas.: Transfer function measurement.}
\end{figure}

This paper aims to present a real-time system for auralizing musical performance on a virtual stage by a single musician or an interactive ensemble between a group of musicians.
The main contribution of this paper is outlining an elaborate framework for interactions between an arbitrary number of musicians on a virtual stage, which is limited only by availability of computing resources and electroacoustic hardware. This includes rigorous considerations of system calibration and latency, which is currently lacking in comparable systems for real-time auralization in literature (e.g., \cite{gari2017}.) 
A proof of concept with two interconnected musicians on a virtual stage with several acoustic variations is also presented. 
The system is primarily intended for studying stage acoustics in virtual settings wherein it is easier to vary the acoustics than in real rooms. 
However, another potential uses include enabling rehearsal/performance on virtual stages.   

Fig.~\ref{fig:auralization_scheme} shows the workflow used in this paper for realizing a virtual acoustics scenario.
The first step is the experimental design for stage acoustics experiments or for another related purpose (e.g. rehearsal session, recording session, training session, etc). 
This design includes the definition of both real and virtual rooms including number, locations, and orientations of both real and virtual sources and receivers as well as microphones and headphones. 
Once the experimental design is set, the second block of activities is the laboratory setup which includes both a physical setup and a digital setup. 
The physical setup involves placing and connecting all the equipment. 
The digital setup includes both the modeling (or measurement) of the binaural room impulse responses (BRIRs) for each virtual source\hyp{}receiver combination and the setup of a Digital Audio Workstation (DAW). 

The third block includes the measurements of latencies, microphone sensitivities, and headphones transfer functions, shown in steps 3.a, 3.b, and 3.c, respectively. 
The step 3c / 4.a involving measurement of headphone sensitivity can be considered as part of both blocks 3 and 4 because it is both a calibration measurement and a part of the activities with participants. 
The fourth block is called "activities with participants" and its main part is the listening/playing tests labeled 4.c (Run test). 
Before running the test, the calibration and latency compensation is applied in step 4.b.
Although the step 4.b does not strictly require the participants to be present, it is grouped in block 4 because the adaption can be carried out quickly while participants remain in the laboratory after step 4.a, which may include time for briefing and completing some general questions forms, as was done in the current case. 
As part of step 4.b, the adapted BRIRs are uploaded to the previously set up  DAW session.
Finally, step 4.c consists of the main test. 

The rest of the paper is organized as follows. Sec.~\ref{sec:setup} introduces the experimental design (block 1) as well as the laboratory setup (block 2) and the BRIR adaption (block 4b).
Sec.~\ref{sec:LatCal} describes the required adaption of the BRIRs in terms of block 3 (namely latency compensation and calibration filters), and also an analysis of the main sources of uncertainty. Finally, in Sec.~\ref{sec:Proof_of_concepts},
a proof of concept is carried out with six guitar duets, followed by the discussion and conclusions in Secs.~\ref{sec:discussion}~and~\ref{sec:conclusions}, respectively.

\section{Setup}
\label{sec:setup}

\begin{figure}[htbp]
	\centering
		\includegraphics[width=0.45\textwidth]{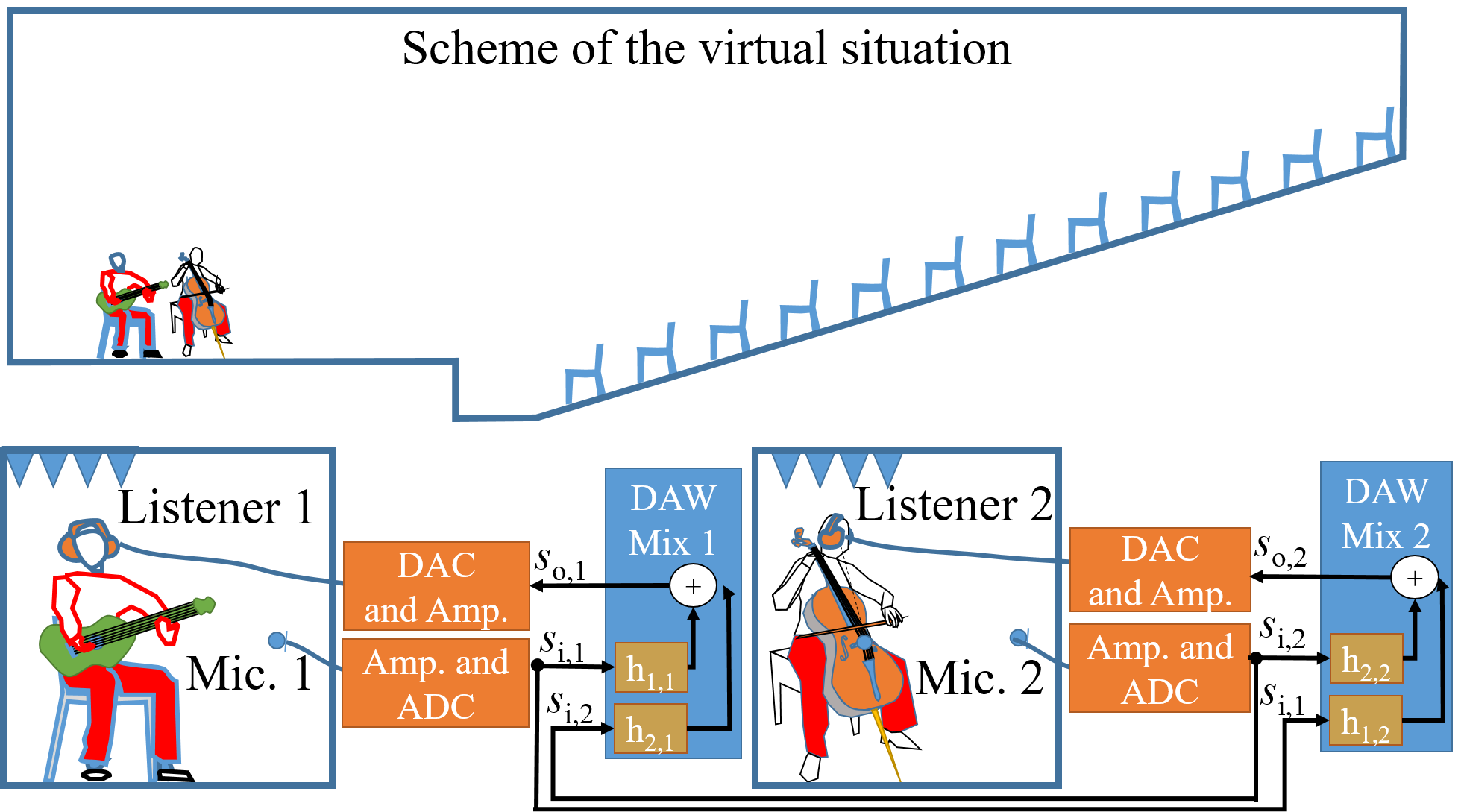}
	\caption{Example of an experimental design scheme.  $s_{\text{i},m}$: input signal from player role $m$,
 $s_{\text{o},n}$: output signal for listener  role $n$, 
 $h_{m,n}$: impulse response for listener $n$ due to player $m$,
  $m\in \{1,2\}$,
  $n\in \{1,2\}$}
	\label{fig:real_imag}
\end{figure}

The goal here is to simulate a virtual stage environment for a solo musician, or for two or more musicians playing simultaneously while each of them is located in an individual isolated booth. 
To that end, each musician hears both the instrument of the other musicians and his or her own instrument in a simulated room in real-time. For simplicity, each musician is defined with a listener and a player role. Furthermore, each musician is assumed to play a musical instrument that can be considered as one source (Fig.~\ref{fig:real_imag}).

Fig. \ref{fig:real_imag} shows an example of the experimental setup (see block 2 in Fig.~\ref{fig:auralization_scheme}) for a system with two players 
and two listeners
, i.e. a system for two musicians in which each of them play one sound source. 
The upper part represents the virtual situation in which a guitarist and a cellist play music in a simulated concert hall. 
The lower part shows the electroacoustic setup (e.g. headphones, microphones, and processing blocks) where the core parts are the DAW mixes which perform the real-time convolutions and the signal summations. 

Let $n\in \{1,\cdots,N\}$ denote the $n$th listener and $m\in \{1,\cdots,M\}$ denote the $m$th player. 
Hence, based on the digital input signal $\textbf{s}_{\text{i},m}$ from each player $m$ and the binaural room impulse response $\textbf{h}_{m,n}$ at the listener $n$ due to each player $m$, the output signal $\textbf{s}_{\text{o},n}$ to be sent to the listener $n$ when $M$ players are simultaneously playing is computed as 
\begin{equation}
\textbf{s}_{\text{o},n} = 
  \textbf{h}_{1,n} \ast \textbf{s}_{\text{i},1} 
+ \cdots 
+ \textbf{h}_{m,n} \ast \textbf{s}_{\text{i},m} 
+ \cdots 
+ \textbf{h}_{M,n} \ast \textbf{s}_{\text{i},M} 	
\label{eq:soj}
\end{equation}
where $\ast$ is the convolution operation. This equation can be straightforwardly implemented in a DAW session. For example, the output signal $\textbf{s}_{\text{o},n}$ for each listener role can be generated by routing $M$ stereo subgroups to a mix channel $n$. Each subgroup $(m,n)$ contains a convolution plug-in which applies the convolution of each of the two channels of $\textbf{h}_{m,n}$ with the monaural signal $\textbf{s}_{\text{i},m}$ routed from the source input channel $m$. Setting up a system for implementing this equation is step 2.b in Fig.~\ref{fig:auralization_scheme}.

The system should be calibrated and latency\hyp{}compensated in order to avoid the effect of the system itself in the signal that arrives at the musicians' ears. 
In this paper, the calibration $\textbf{k}_{m,n}$ and the latency-compensation due to both the audio system latency $t_\text{l}$, and the physical source\hyp{}microphone distance $d_{\text{M-S}}$ are applied to the simulated BRIRs $\textbf{h}'_{m,n}$ as

\begin{equation}
    \textbf{h}_{m,n} = \textbf{h}'_{m,n} \ast \sinc(\textbf{t}_\text{s}-t_\text{l}-\frac{{d}_{\text{M-S},m}}{c}) \ast \textbf{k}_{m,n} 
    \label{eq:_calib_uncal}
\end{equation}
where $\textbf{t}_\text{s}=[0, 1/f_\text{s}, \cdots, T]$ is the time vector with sampling frequency $f_\text{s}$ and $c$ is the speed of sound. The calibration filter $\textbf{k}_{m,n} $, the latency compensation $t_\text{l}$ and distance compensation $d_{\text{M-S}}/c$ are analyzed in detail in the next section. 
This equation is applied in step 4.b in Fig.~\ref{fig:auralization_scheme}. Although the result of eq. \eqref{eq:_calib_uncal} is required in eq. \eqref{eq:soj}, it is simpler to first set up the implementation of eq. \eqref{eq:soj} in a DAW and then just adapting the BRIRs $\textbf{h}_{m,n}$ for each participating musician according to eq. \eqref{eq:_calib_uncal}.
In most cases the experimental design involves fixed gains of the preamplifiers and the distance between microphone and musical instrument; hence, the only element of the calibration $\textbf{k}_{m,n}$ that is modified for each participant is the headphone transfer function. 

Due to latency issues (Sec.~\ref{sec:latency}), the direct sound is skipped in the impulse response $\textbf{h}_{m,n}$ with $m=n$ (listener $n$ receives the direct sound from his or her own instrument), and is included in the impulse response $\textbf{h}_{m,n}$ with $m\neq n$ (because listener $n$ is not in the same room than player $m$). 
For example, in Fig. \ref{fig:real_imag}, the direct sound of cello is skipped in $\textbf{h}_{1,1}$ and included in $\textbf{h}_{2,1}$.

In the example of Fig. \ref{fig:real_imag}, the guitarist is the listener $n=1$ and hears the  signal $\textbf{s}_{\text{o},1} = \textbf{h}_{1,1} \ast \textbf{s}_{\text{i},1} +   \textbf{h}_{2,1} \ast \textbf{s}_{\text{i},2}$ which is processed in real time in the mix 1 of the DAW according to Eq. \eqref{eq:soj}. 
For that reason, the signals $\textbf{s}_{\text{i},1}$ and $\textbf{s}_{\text{i},2}$ are picked using microphones in front of each instrument for the guitar and the cello, respectively. 
Simultaneously, the mix 2 of the DAW renders audio for the cellist who represents the listener $n~=~2$. 
For that purpose, the mix 2 of the DAW implements Eq. \eqref{eq:soj} by processing the same $\textbf{s}_{\text{i},1}$ and $\textbf{s}_{\text{i},2}$ signals, but this time convolved with $\textbf{h}_{1,2}$ and $\textbf{h}_{2,2}$, respectively (i.e. ${{\textbf{s}_{\text{o},2} = \textbf{h}_{1,2} \ast \textbf{s}_{\text{i},1} +   \textbf{h}_{2,2} \ast \textbf{s}_{\text{i},2}}}$). 

The considered calibration accounts for both conversions from acoustic signals to digital signals and from digital signals to acoustic signals. Sec. \ref{sec:Calibration} presents a method for estimating the calibration filters and the analysis of the effects of errors in the input data for this estimation. 

\section{Latency compensation and system calibration}
\label{sec:LatCal}

The calibration of the virtual acoustic system can be calculated as the frequency-dependent gain for sending the signal at the microphone in front of player $m$ to the headphones of listener $n$. This represent the signal that would be received at the entrance of his or her ear canal in a real situation in free field without the headphones.  The calibration filter is defined in the frequency domain of the signals as $K_{m,n}(f)$.

\begin{figure}[ht]
	\centering
		\includegraphics[width=0.45\textwidth]{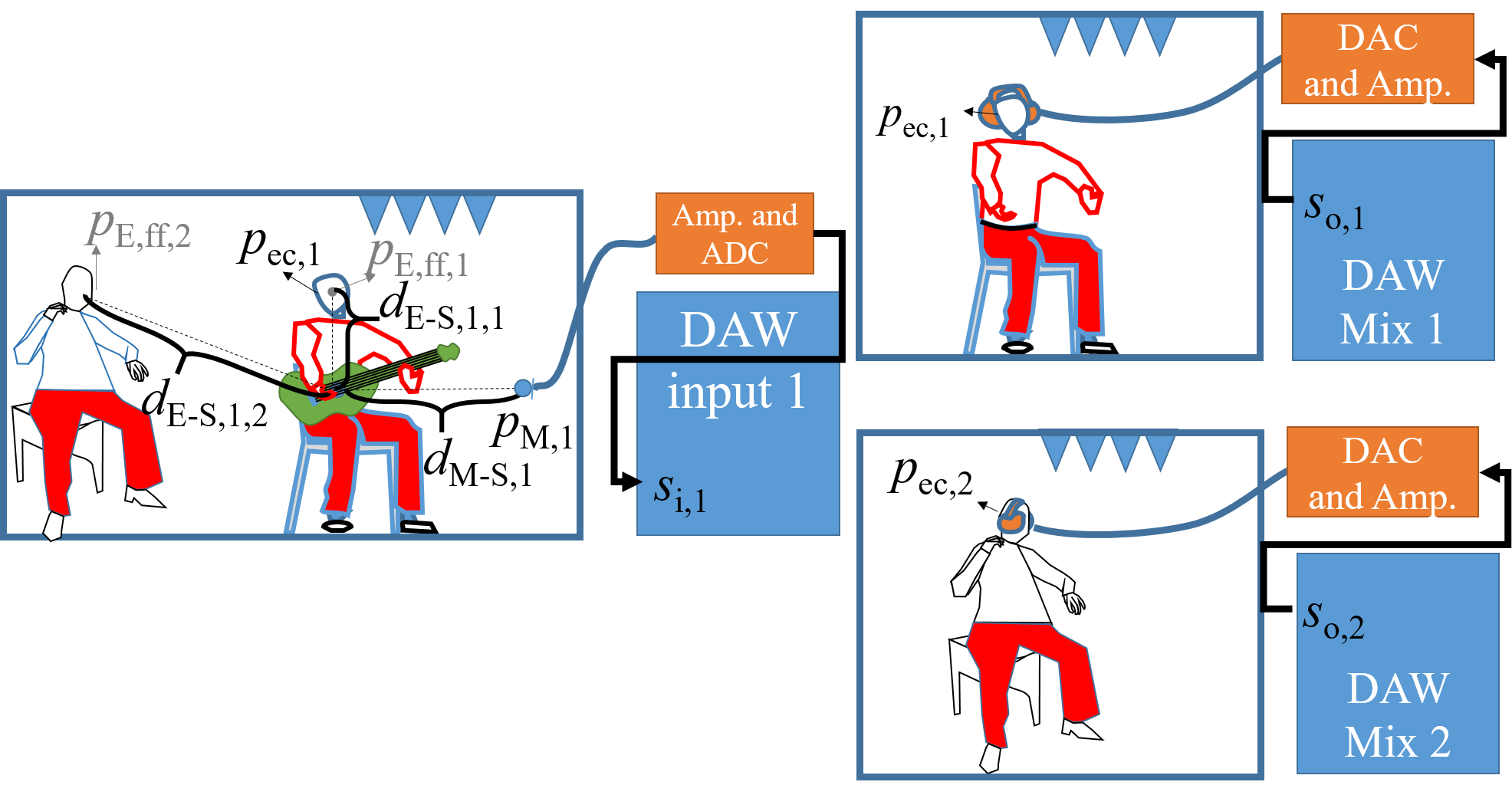}
	\caption{Scheme of calibration factors required for real-time auralization. Left part: player role. right part: listener roles.
$p_{\text{ec},n}$ sound pressure at the entrance of the ear canal of listener $n$, 
   ${p}_{\text{E,ff},n}$: free-field sound pressure at the position of the head of listener $n$,
 $d_{\text{M-S},m}$ distance from the acoustic center of source $m$ to microphone $m$,
  $d_{\text{E-S},m,n}$ distance from the acoustic center of source $m$ to the head of player $n$, 
   ${p}_{\text{M},m}$: sound pressure at a microphone in front of player $m$,
 $s_{\text{i},m}$: input signal from player role $m$,
 $s_{\text{o},n}$: output signal for listener  role $n$, 
  $m\in \{1\}$,
  $n\in \{1,2\}$
 }
	\label{fig:guit}
\end{figure}

Fig.~\ref{fig:guit} shows a scheme of the proposed situation for deducing the calibration filters in a hypothetical anechoic environment. 
The left part in Fig.~\ref{fig:guit} shows the hypothetical real situation which represents the playing situation with $N=2$ listeners, but isolated in an anechoic environment. 
In this example, just $M=1$ player  plays a guitar, so the input signal is just $s_{\text{i},1}$ picked at a microphone placed in front of the guitar. 
The right part in Fig.~\ref{fig:guit} shows the auralization situation.

Let $\hat{s}(f)$, $\hat{p}(f)$, and $\hat{h}(f)$ be the elements of frequency $f$ of the Fourier transforms of digital signals $\textbf{s}$, sound pressure signals $\textbf{p}$, and impulse responses $\textbf{h}$, respectively. 
In the case of Fig. \ref{fig:guit} the right side of Eq. \eqref{eq:soj} keeps just one term corresponding to the only musical instrument in the proposed situation. 
The element of frequency $f$ of the Fourier transform of Eq.~\eqref{eq:soj} is  
\begin{equation}
\hat{s}_{\text{o},n}(f) =  \hat{h}_{\text{ane},m,n}(f) 
\hat{s}_{\text{i},m}(f) , 
\label{ykx}
\end{equation}
 with latency compensated and frequency calibrated impulse response 
\[\hat{h}_{\text{ane},m,n}(f) = \hat{h}'_{\text{ane},m,n}(f) K_{m,n}(f) e^{-j\omega (t_\text{l}+\frac{\tilde{d}_{\text{M-S},m}}{c})}\]
as an alternative of Eq. \eqref{eq:_calib_uncal}. 
The notation below will drop $f$ for brevity. 

\subsection{Latency}
\label{sec:latency}

The latency $t_\text{l}$ of off-the-shelf sound interfaces is usually greater than the wave travel-time for listening to a musical instrument played by oneself (e.g. shorter than 0.2~ms for a violinist or about 1.5~ms for a guitarist; some exceptions include pipe-organs, bagpipes, and some percussion instruments). 

This is demonstrated in Table~\ref{tab:laten}, which shows the latency of three interfaces measured using the ITA Toolbox \cite{ITA-Toolbox_2017}, 
and where each interface was set to its smallest buffer size option. 
\begin{table}[ht]
\caption{Measured latency of RME Fireface UC, FOCUSRITE Scarlett 2i2, and M-AUDIO Fast Track Ultra interfaces \label{tab:laten}}
\begin{center}
\begin{tabular}{l r r r}
Interface & $S_\text{B}$ (\text{samples}) & $t_\text{l}$ (ms) & $e_\mathrm{d}$ (m)\\
\hline
Fireface UC& 48 & 4 & 1 \\
Scarlett 2i2& 48 & 13 & 4 \\
Fast Track Ultra& 256 & 17 & 6 \\
\end{tabular}
\end{center}
\end{table}

To overcome the hardware latency issue for simulating hearing one's own instrument, one approach includes latency compensation, i.e. time shifting the impulse response. This may be applied without cropping the impulse response for source\hyp{}receiver distances no closer than an equivalent distance $e_\mathrm{d} = c \times t_\text{l}$. Another approach, which is also used in previous systems (e.g., \cite{,p-g_equal_2011,y_c_m_2012}), includes skipping the simulation of the direct sound for the hearing one's own instrument. This approach involves allowing the actual direct sound to reach the musician through open headphones. 

Hence, once the direct sound for hearing oneself is skipped, the compensation time limit is the propagation time due to distances between musicians for hearing others, or due to floor reflection path for hearing oneself. For the current implementation, the RME Fireface UC interface meets both these limits: for distances greater than 2~m between musicians and for receiver heights greater than 1.3~m (these estimations include a compensation for a microphone\hyp{}source distance of $d_{\text{M-S}}=1$~m), and represents a suitable off-the-shelf interface (Table~\ref{tab:laten}).

	\subsection{Calibration}
\label{sec:Calibration}

During auralization of the proposed situation (Fig. \ref{fig:guit}) with headphones, listeners receive the same sound pressure at the ear canal $p_{\text{ec},n}$ that they would receive in an anechoic environment due to her or his own musical instrument (as in the right-upper part of Fig. \ref{fig:guit}) or the instrument of another musician (as in the right-lower part of Fig. \ref{fig:guit}).
To that end, the output signal $s_{\text{o},n}$ should contain the required calibration to yield a sound pressure $p_{\text{ec},n}$ at the entrance of the ear canal of each musician. 

\subsubsection{Filter components}
\label{sec:flit_comp}

In this section, the filter components of Eq. \eqref{ykx} are discussed and estimates of the filter functions are derived. For this purpose, estimates of the sound pressure at the microphone in front of each player and at the ears of each listener are defined first. 

\begin{figure}[ht]
	\centering
		\includegraphics[width=0.38\textwidth]{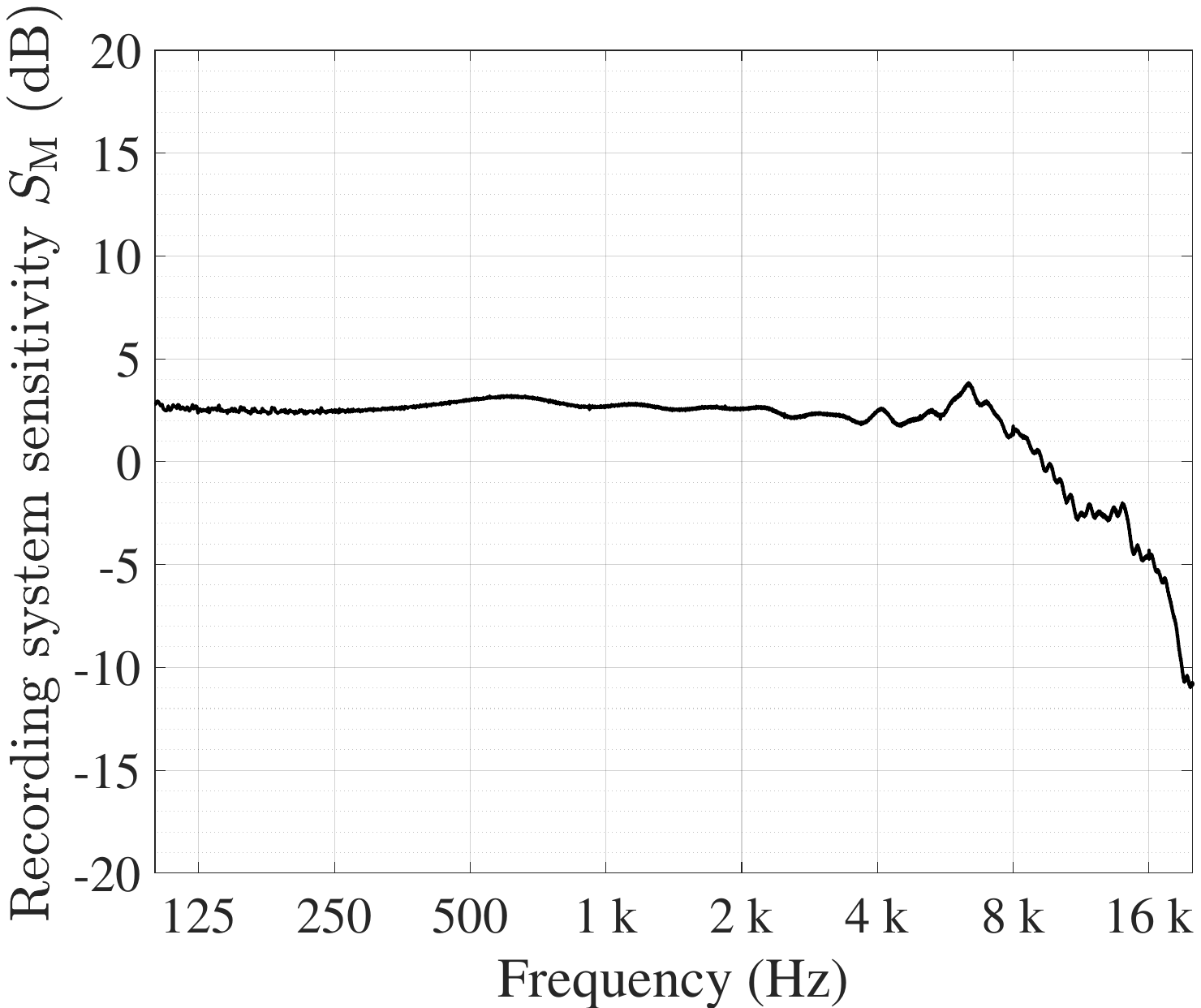}
	\caption{Recording system sensitivity $S_{\text{M},m}$ (dB) of Oktava MK-012 microphone with RME FIREFACE UC interface.}
	\label{fig:SM}
\end{figure}

The sound pressure at the microphone $p_{\text{M},m}$ can be obtained from the input signal $s_{\text{i},m}$ as
\begin{equation}
\hat{p}_{\text{M},m} = \frac{\hat{s}_{\text{i},m}}{S_{\text{M},m}}
\label{eq:p_sSM}
\end{equation}
where $S_{\text{M},m}$ is the digital sensitivity of the recording system measured as the ratio of the digital signal caused by 1 Pa of sound pressure at the microphone to the digital full scale of the system (i.e. it includes microphone, amplifier, ADC). 
Fig. \ref{fig:SM} shows the digital sensitivity of the recording system consisting of an Oktava MK-012 microphone with an RME FIREFACE UC interface which is used below for a proof of concept (section IV). 
It was measured by comparison to a calibrated class 1 measurement microphone in the center of a medium-sized room with reverberation time below $0.1\ \text{s}$.
Reflections were croped with a 15-ms-length rectangular time-window centered in the direct sound as proposed in \cite{accolti2021}. 

Let $W_m$ be the sound power for source $m$, $Q(\Omega)$ be the directivity factor for direction $\Omega$, and $\Omega_{\text{M-S},m}$ be the microphone--source direction for source $m$. 
Then, the free-field squared sound pressure $\hat{p}^2_\text{M}$ measured by a microphone located at a distance $d_{\text{M-S},m}$ from the acoustic center of source $m$ is
\begin{equation}
\hat{p}^2_{\text{M},m} = \frac{\rho c W_m Q(\Omega_{\text{M-S},m})}{4 \pi d^2_{\text{M-S},m}} e^{2 \pi f \frac{d_{\text{M-S},m}}{c}}
\label{eq:LMd}
\end{equation}
where $\rho$ is the air density and $c$ is the speed of sound.
Note that the location of the acoustic center of the source depends on frequency \cite{hagai_acoustic_2011}. Hence, $W_m$, $Q(\Omega_{\text{M-S},m})$, and also $d_{\text{M-S},m}$ are functions of frequency. 

Let $\Omega_{\text{E-S},m,n}$ be the direction and $d_{\text{E-S},m,n}$ the distance from source $m$ to listener $n$. 
Then, the free-field squared sound pressure at the position of the head of listener $n$ but in absence of the listener is 
\begin{equation}
\hat{p}^2_{\text{E,ff},n} =  \frac{\rho c W_m Q(\Omega_{\text{E-S},m,n})}{4 \pi d^2_{\text{E-S},m,n}}
\label{eq:LHd}
\end{equation}

\begin{figure}[ht]
	\centering
		\includegraphics[width=0.38\textwidth]{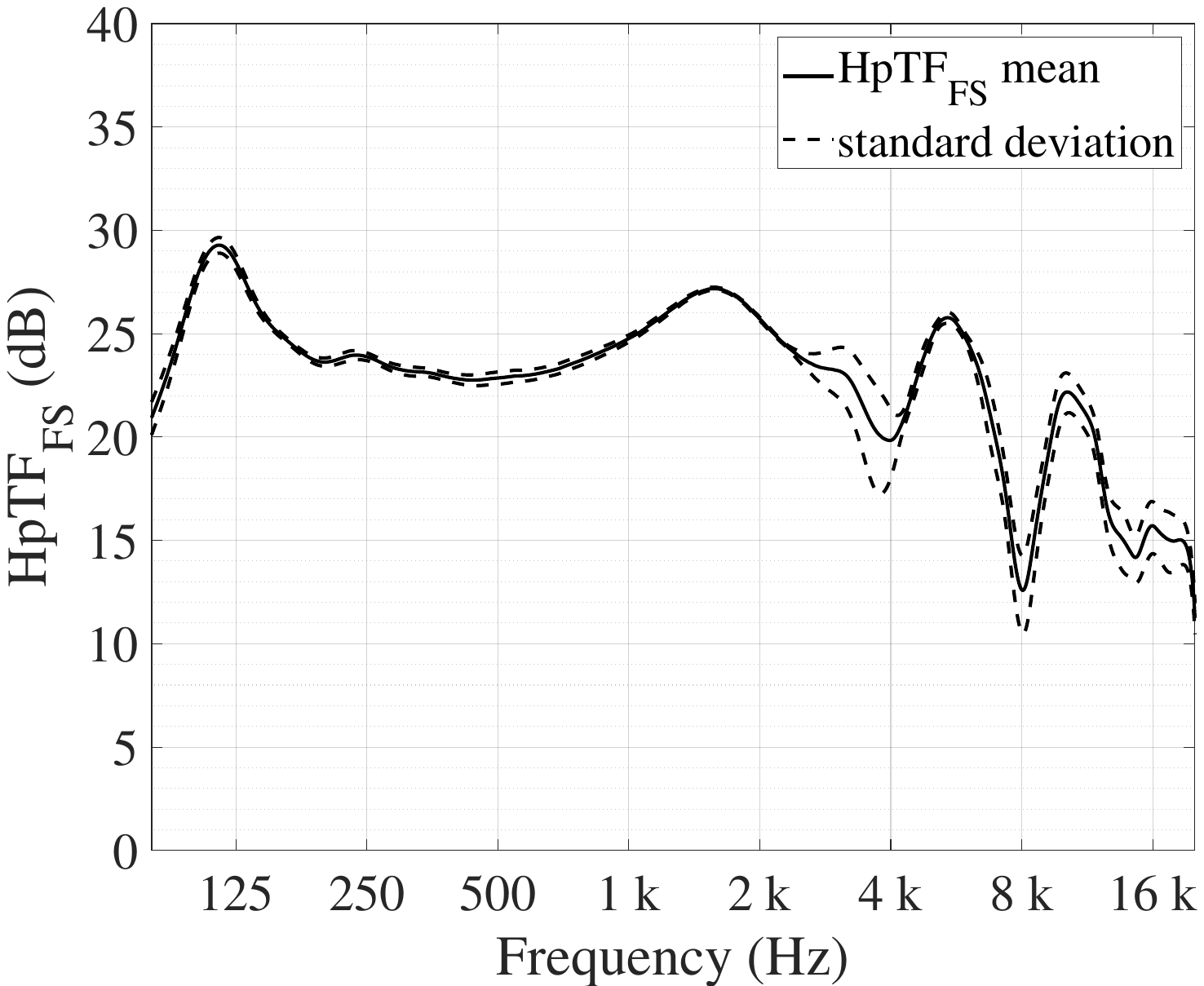}
	\caption{Headphone transfer function $H_{\text{E},n}$ (dB) of LAMBDA STAX headphones with RME FIREFACE UC interface.}
	\label{fig:SE}
\end{figure}

On the one hand, the nominal sound pressure at the blocked entrance of the ear-canal $\hat{p}_{\text{ec},n}$ is 
\begin{equation}
\hat{p}_{\text{ec},n} = \hat{p}_\text{E,ff,n} H_{\text{S},m,n} 
\label{eq:LHs}
\end{equation}
where $H_{\text{S},m,n}$ is the head related transfer function (HRTF) of listener $n$ in the direction to the source $m$.
On the other, the actual sound pressure at the blocked entrance of the ear canal $\hat{p}_{\text{ec},n}$ can be obtained from an output signal $\hat{s}_{\text{o},n}$ as
\begin{equation}
\hat{p}_{\text{ec},n} = \hat{s}_{\text{o},n} H_{\text{E},n} e^{j2\pi t_\text{l}}
\label{eq:pec_sSE}
\end{equation}
where $H_{\text{E},n}$ is the headphone transfer function (HpTF) of the playback system for the listener $n$ measured in terms of sound pressure with reference to the digital full scale (i.e. it includes the response of headphones, amplifier, and DAC). 
Fig. \ref{fig:SE} shows the modulus in decibels of the HpTF of the playback system consisting of a LAMBDA STAX headphone with an RME FIREFACE UC interface. 
These measurements were carried out with a dummy-head with ten repetitions in order to estimate the deviations related to headphones placement. 
Besides, these results are congruent with results previously obtained by \cite{hammershoi_methods_2002} in blocked ear canals of humans. 

The calibration is based on a situation of direct sound in free field because it avoids propagating errors from estimations of the acoustic properties of room surfaces which would be the case for calibration based on reflections.   
Although, the actual direct sound for hearing oneself passes through the open headphone during the auralizations, the calibration is valid for the simulated impulse response that skips direct sound. 
Continuing with headphones properties, the sound insulation is important when skipping the direct sound of simulations because the actual direct sound is expected to reach ears through them (Subsec. \ref{subsec:calforguit}). 

Let $\ell=e^{-2\pi f ({(}d_{\text{M-s},m}/{c}+t_\text{l})}$ be the latency compensation. Then, substituting Eqs. \eqref{eq:p_sSM}, \eqref{eq:LMd}, \eqref{eq:LHd}, and \eqref{eq:LHs} in \eqref{eq:pec_sSE} leads to 
\begin{equation}
\hat{s}_{\text{o},n} =\frac{ d_{\text{M-S},m} \sqrt{Q(\Omega_{\text{E-S},m,n})} }{H_{\text{E},n} S_{\text{M},m} d_{\text{E-S},m,n} \sqrt{Q(\Omega_{\text{M-S},m})}} H_{\text{S},m,n} 
\ell
\hat{s}_{\text{i},m} 
\label{eq:LHLM_prev}
\end{equation}
where directivity factor $Q$ is referred to the mean squared pressure $\overline{p^2}$ across directions $\Omega$ over a spherical surface centered at the source, 
i.e. 
\begin{equation}
Q(\Omega) = \frac{p^2(\Omega)}{\overline{p^2}}
\label{eq:Q}
\end{equation}
However, the directional factor
\begin{equation}
\Gamma(\Omega)=\frac{p(\Omega)}{p(\Omega_\text{ref})}
\label{eq:Gamma}
\end{equation}
is usually preferred as input-data for geometrical acoustics (GA) software. 
Hence, substituting eq.~\eqref{eq:Q} and eq.~\eqref{eq:Gamma} in eq.~\eqref{eq:LHLM_prev}, yields to 
\begin{equation}
\hat{s}_{\text{o},n} =\frac{ d_{\text{M-S},m} \Gamma(\Omega_{\text{E-S},m,n}) }{H_{\text{E},n} S_{\text{M},m} d_{\text{E-S},m,n} \Gamma(\Omega_{\text{M-S},m})} H_{\text{S},m,n} 
\ell
\hat{s}_{\text{i},m}
\label{eq:LHLM}
\end{equation}
Furthermore, since the reference direction coincides with the position of the microphone in front of the source, the respective sound pressures are equivalent, i.e. $p(\Omega_{\text{M-S},m})=p(\Omega_\text{ref})$. 
Thus, the directional factor in the source\hyp{}microphone direction is unity, i.e. $\Gamma(\Omega_{\text{M-S},m}) = 1$.

The distance parameters $d_{\text{M-S},m}$ and $d_{\text{E-S},m,n}$ should be measured from the acoustical center of the source $m$ to the acoustical center of the microphone in front of that source and to the head of listener $n$, respectively. 
Hence, all the parameters in Eq. \eqref{eq:LHLM} are frequency-dependent, because the acoustical center of a complex source such as a musical instrument is different for each frequency. 
However, source\hyp{}receiver distance is usually a single value parameter in the state-of-the-art GA simulators \cite{savioja2015} which can be used to estimate the impulse responses $h_{m,n}$. 
In order to have an estimate of this effect in the calibration, in this work it is assumed 
\begin{equation}
d_{\text{M-S},m}/d_{\text{E-S},m,n}= \tilde{d}_{\text{M-S},m}/\tilde{d}_{\text{E-S},m,n} E_{\text{K},m,n}
\label{eq:d_E}
\end{equation}
where $\tilde{d}_{\text{M-S},m}$ and $\tilde{d}_{\text{E-S},m,n}$ are single value estimations of $d_{\text{M-S},m}$ and $d_{\text{E-S},m,n}$, respectively. Besides, $E_{\text{K},m,n}$ is introduced as the error due to these two estimations (subsec. \ref{exAnexA} shows a detailed description of the effect of this source of error on the calibration). 

Eq. \eqref{eq:LHLM} can be written as Eq. \eqref{ykx} (i.e. $\hat{s}_{\text{o},n} =  \hat{h}_{\text{ane},m,n} \hat{s}_{\text{i},n}$ and $\hat{h}_{\text{ane},m,n} = \hat{h}'_{\text{ane},m,n} K_{m,n} \ell$) with 
\begin{equation}
K_{m,n} =\frac{\tilde{d}_{\text{M-S},m} E_{\text{K},m,n}} {H_{\text{E},n} S_{\text{M},m} \Gamma(\Omega_{\text{M-S},m})}
\label{eq:k}
\end{equation}
and
\begin{equation}
\hat{h}'_{\text{ane},m,n} = \frac{\Gamma(\Omega_{\text{E-S},m,n}) H_{\text{S},m,n} } {\tilde{d}_{\text{E-S},m,n}}
\label{eq:hat_h}
\end{equation}

Note that Eq. \eqref{eq:k} can be used for estimating the calibration $K_{m,n}$ for hearing oneself (i.e. $m=n$) as well as for hearing others (i.e. $m\neq n$). Besides, the calibration for the same listener $n$ for each source $m$ would modify the values of all parameters in Eq. \eqref{eq:k} except $H_{\text{E},n}$ that holds the same for hearing oneself as well as for hearing others. 

In auralizations, Eq. \eqref{ykx} also applies but the free-field transfer function $\hat{h}_{\text{ane},m,n}$ which was assumed for deducing the calibration data is replaced by the room transfer function $\hat{h}_{m,n}$. 
Hence, the head related transfer function $H_\text{S}$, the source-head distance $\tilde{d}_\text{E-S}$, and source directional factor $\Gamma(\Omega_\text{E-S})$ may be applied to the binaural room impulse responses (BRIRs) which is the state-of-the-art of GA simulators. In this article the BRIRs are calculated with RAVEN \cite{schroder_raven_2011} and the input data include the source directivity data reported in \cite{weinzierl_database_2017}.

\subsubsection{The effect of source\hyp{}receiver distance} 
\label{exAnexA}

The calibrated output signal depends on two distances measured from the same source position as shown above in Eq. \eqref{eq:LHLM}.
One of these distances is measured to the head of a listener (i.e. $d_\text{E-S}$) and the other one to the center of a microphone (i.e. $d_\text{M-S}$). 

Let $\tilde{d}_\text{M-S}$ and $\tilde{d}_\text{E-S}$ be estimations of $d_\text{M-S}$ and $d_\text{E-S}$, respectively.  
These  estimations -- which assume the acoustic center is placed in a fixed point for all frequencies -- can be the geometric center, the mean of the acoustic center for all frequencies in which it was aligned for the directivity database in \cite{shabtai_acoustic_2015}, or any other suitable approximation. 
Then, the errors $e_1$ and $e_2$ indicate the estimation error for ${d}_\text{M-S}$ and ${d}_\text{E-S}$, respectively, and the ratio of this two distances is

\begin{equation}
\frac{d_\text{M-S}}{d_\text{E-S}} = \frac{\tilde{d}_\text{M-S}+e_1}{\tilde{d}_\text{E-S}+e_2} 
\label{eq:dd}
\end{equation}

\begin{figure}[ht]
	\centering
		\includegraphics[width=.30\textwidth]{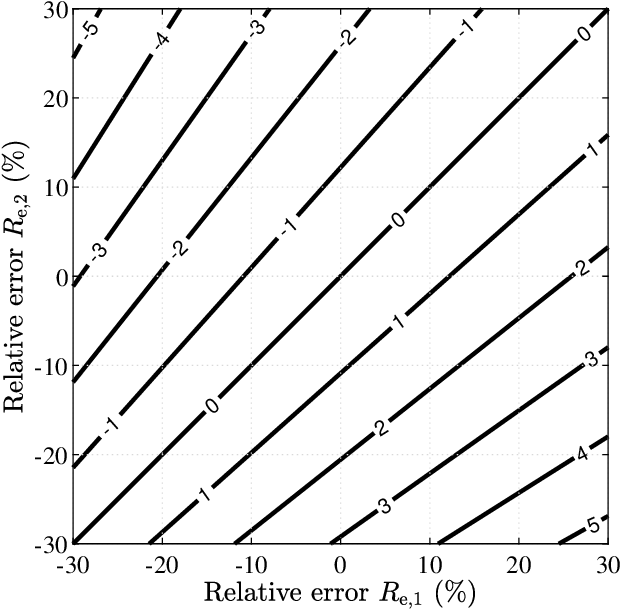}
	\caption{Effect of distances estimation error in dB units ($E_\text{K}$) on calibration}
	\label{fig:error}
\end{figure}

In order to study the effect of the distance estimation error on the calibration $K_{m,n}$, 
eq. \eqref{eq:d_E} was proposed. Hence, inserting eq. \eqref{eq:d_E} in eq. \eqref{eq:dd} gives the effect of the estimation error 
\begin{equation}
E_\text{K} = \frac{1+e_1/\tilde{d}_\text{M-S}}{1+e_2/\tilde{d}_\text{E-S}} 
\label{eq:error}
\end{equation}
Fig. \ref{fig:error} shows results of equation \eqref{eq:error} as contour lines in dB units related to distance relative errors $R_\mathrm{e,1}= e_1/\tilde{d}_\text{M-S}$ and $R_\mathrm{e,2}=e_2/\tilde{d}_\text{E-S}$, .  

The errors $e_1$ and $e_2$ for each frequency depend on the position of the acoustic center of the source and do not depend on the position of the receiver\footnote{Actually the acoustic center of the microphone can vary by about a few centimeters, but it is assumed negligible compared to the variation of the acoustic center of the source in this article} for a given source-receiver direction.
Therefore the relative errors $R_\mathrm{e,1}$ and $R_\mathrm{e,2}$ are smaller for larger source\hyp{}receiver distances ($\tilde{d}_\text{M-S}$ and $\tilde{d}_\text{E-S}$). In other words, the effect of these relative errors in the calibration estimated with single value distances can be smaller for receivers in the audience than for receivers on the stage, because the former are closer to the sources than the latter.

In the case of a violin, the acoustic center for low frequencies can be about 15 cm apart from the acoustic center for high frequencies, as reported in \cite{shabtai2015}. Considering a distance of about 1 m this could yield a relative error $R_\mathrm{e,1} \approx 15\%$. 

The relative error $|R_\mathrm{e,1}|$ can reach about $15\%$ for a cello player hearing himself or herself; thus the magnitude of the effect of the distance estimation error $E_\text{K}$ can reach values below $2.6$ dB for $|R_\mathrm{e,2}| \leq |R_\mathrm{e,1}|$. For singers, woodwind instruments, brass instruments, and other instruments in which the source is very close to the ears, the $|R_\mathrm{e,2}|$ error is likely greater. Hence, the proof of concepts in this paper is carried out with acoustic guitars that are usually played at a considerable distance from the ears.  

For a given relative error $R_\mathrm{e,1}$, the effect is null when $R_\mathrm{e,2}=~R_\mathrm{e,1}$; it can be the case that both receivers are at the same point or a geometric coincidence which are both unlikely to occur in practice. This is shown in the curve for 0 dB effect in Fig. \ref{fig:error}. Furthermore, for $R_\mathrm{e,1}>R_\mathrm{e,2}$ the effect is positive in magnitude, whereas, for $R_\mathrm{e,1}<R_\mathrm{e,2}$ the effect is negative in magnitude. 

\subsubsection{Source directivity approach}
\label{sec:SourceDirectivityApproach}

\begin{figure}[ht]
	\centering
		\includegraphics[width=0.18\textheight]{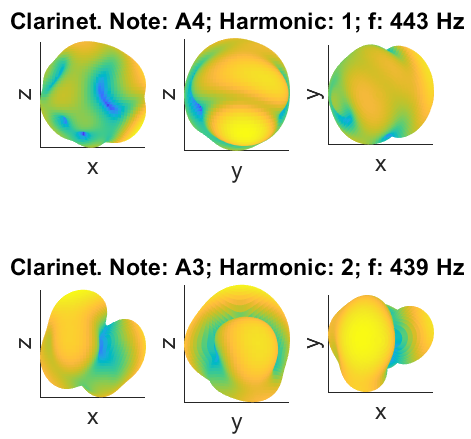}
		\includegraphics[height=0.16\textheight]{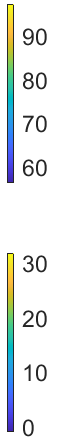}
	\caption{Comparison of balloon plots for a partial tone generated with the low register of a clarinet and a fundamental tone with almost the same frequency  but played in the high register. Legend in dB scale (data extracted from \cite{weinzierl_database_2017}).}
	\label{fig:compare_dirs_clarinet}
\end{figure}

The directivity of musical instruments occurs due to their complex physical systems which involve a large number of variables.  
However, the state-of-the art assumes fractional octave bands as a simplified representation of the radiation patterns because it reduces computational costs of simulations. 
Nevertheless, this simplification may not be totally accurate for some partials in some musical instruments \cite{hohl_2009,hohl_2010,patynen_directivities_2010,accolti2022}. 
Besides, some partials of the same frequency may have strong dissimilarities for certain instruments.
For instance, 
Fig.~\ref{fig:compare_dirs_clarinet} shows balloon plots of a clarinet for the same $440\ \text{Hz}$ frequency due to the fundamental tone of an A4 note in the upper register and due to the first harmonic of a A3 note in the lower register. 

The input data representing directivity can be extracted from several available databases  \cite{otondo_influence_2004,katz_directivity_nodate,hohl_2009,pollow_measuring_2009,patynen_directivities_2010, weinzierl_database_2017,shabtai_generation_2017}.
In this article the directivity database by \cite{weinzierl_database_2017} is chosen because of its high resolution and its centering processing. 
This database provides up to 4-th order spherical harmonics coefficients for 41 musical instruments of different historical periods considering up to 10-th order partials for each note in a chromatic scale covering the typical range of each instrument. Then,  the directivity for each band is estimated by averaging the directional index of all the partials that fall in that band for that instrument \cite[sec. II-C]{accolti2022}.

The directivity for the same frequency and two different partials can be both as similar as about $0.1$ dB for certain directions and as different as about $40\ \text{dB}$ for certain other directions 
However, it is shown below that the relative likelihood for such huge discrepancies is small compared to that for similarities. 

\begin{figure}[ht]
	\centering
	\includegraphics[width=.45\textwidth]{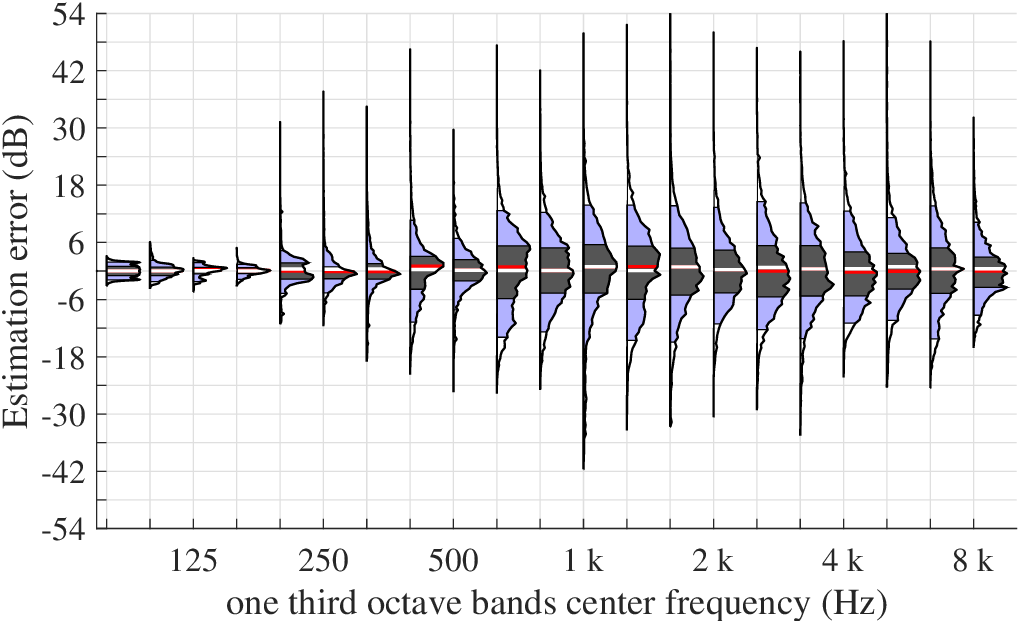}
	\caption{Estimated probability density functions of directional factor estimation errors for a guitar. The dashed area shows the percentile 25 and 75, the red solid line shows the mean and the white solid line shows the median.}
	\label{fig:Q_estimation_errors}	
\end{figure}

Fig.~\ref{fig:Q_estimation_errors} shows the probability density function (PDF) of the estimation errors of the directivity for each third octave frequency band for a guitar \cite{accolti2022}. 
These estimation errors are defined in each direction as the level differences between the actual directional index for each partial and the average of the the directional index for the corresponding third octave band. 
Then, the errors are evaluated at $2,592$ directions over a grid of azimuth and elevation with a 5° resolution. Finally, the PDFs are estimated for each frequency band with the kernel method\footnote{Gaussian 0.1 dB Kernel} on the errors for each of the partials that fall in that band. 

The error of the $50\ \%$ of the evaluated directions is at most about 6 dB, as shown in Fig.~\ref{fig:Q_estimation_errors}. It should be noted that the error for partials is likely less noticeable than for fundamental tones, because the latter usually carry more energy than higher order partials. As shown, errors are (surprisingly) large but this is what is currently possible with the state-of-the-art databases.

Musicians movements may also cause an effect on the radiation pattern as well as the source position and orientation. Thus, the movements of musicians can be incorporated as in \cite{ackermann_2019} by the RAVEN's animation module. However, in this article the sound field is simulated just with static sources as a proof of concept.

\section{Proof of concept}
\label{sec:Proof_of_concepts}

In this section, the implementation and validation of a two-players – two-listeners setup in an  virtual-acoustics interactive system is described. To that end, the three subsections show an analysis comparing a real measurement with a  simulation, an objective determination of the calibration functions, and a subjective evaluation of the system, respectively.

\subsection{Simulation of a measured setup}
\label{sec:SimulationOfAMeasuredSetup}

\begin{figure}[ht]
\begin{center}
\includegraphics[width=0.25\textwidth]{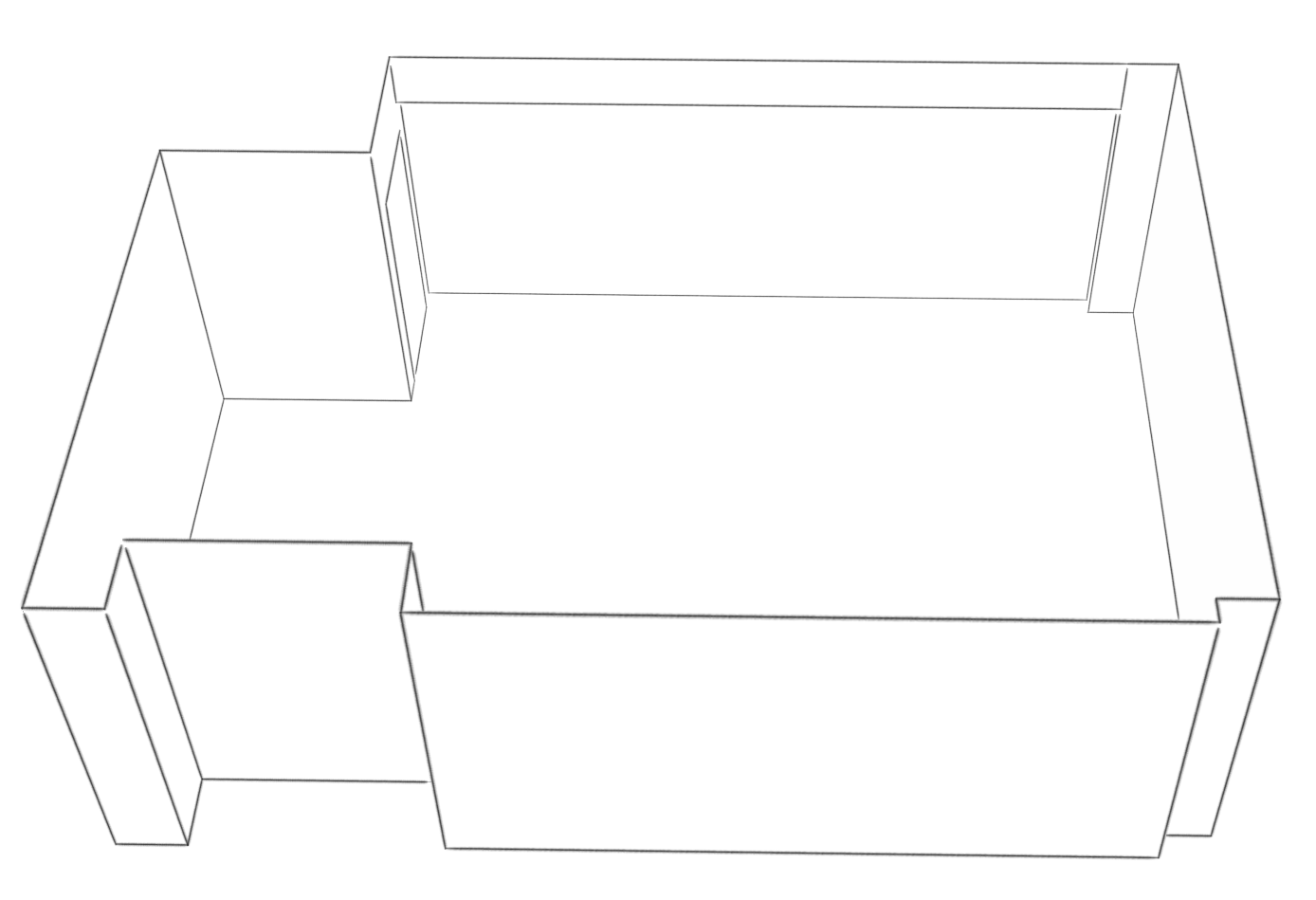}%
\end{center}
\caption{Room geometry}%
\label{Raven_room}%
\end{figure}

In this subsection a comparison of simulation versus measurement of a setup for one musician hearing oneself is documented. Therefore, a real room with a dummy head and a loudspeaker was both measured and simulated.
The room is about $6\ \text{m}$ wide, $8\ \text{m}$ long, and $3\ \text{m}$ high;  
the inner surfaces are acoustically hard with mean absorption coefficient of $0.06$ \cite[See Scene 9 in][]{BRAS2019}. 
Fig. \ref{Raven_room} shows the room geometry. 

\begin{figure}[h!]
\begin{center}
\includegraphics[width=0.30\textwidth]{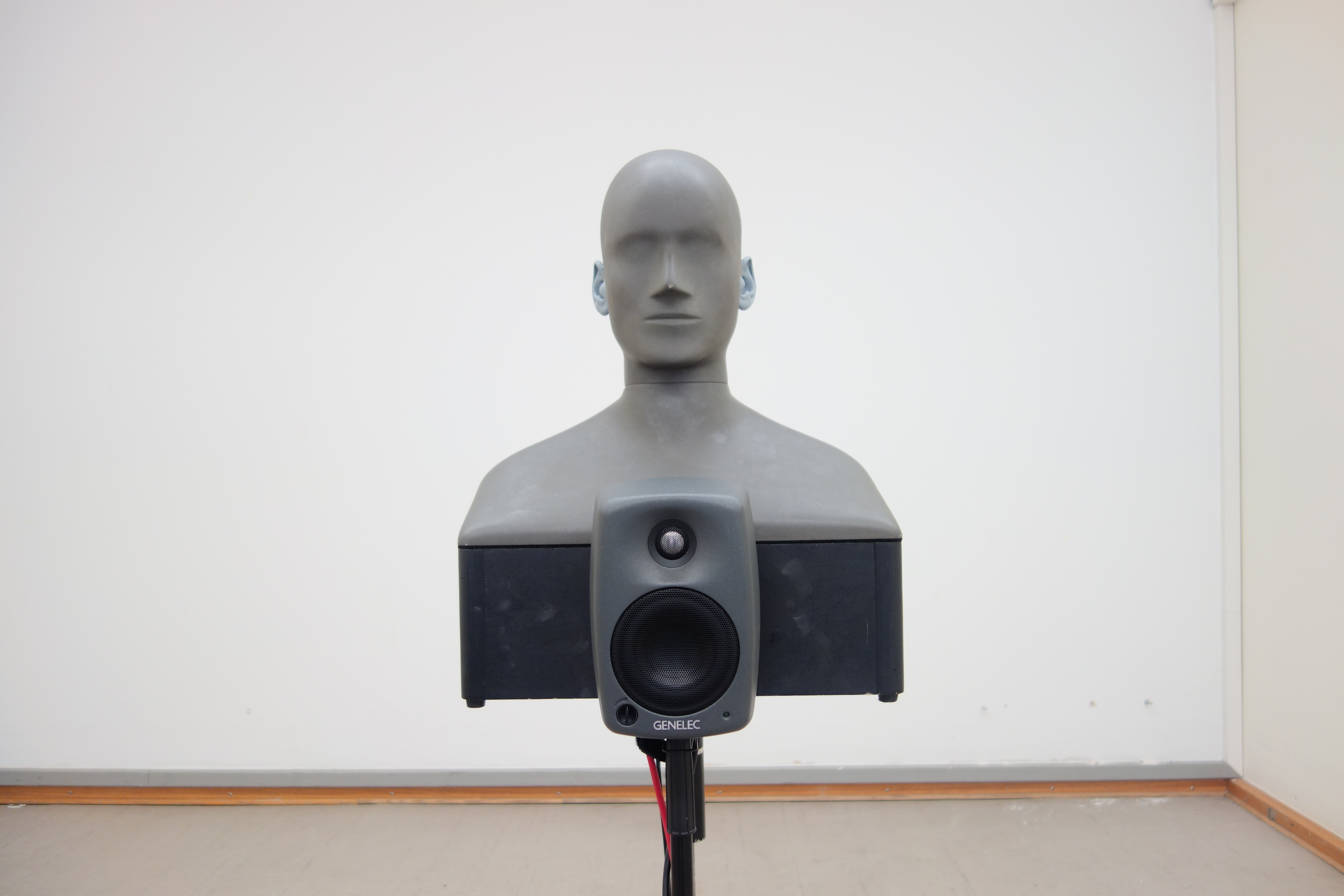}%
\end{center}
\caption{Measurement setup for binaural room impulse response using a Genelec 8020c loudspeaker and a dummy head in a $146 \text{ m}^3$ room}%
\label{BRIRsetup}%
\end{figure}

A binaural room impulse response (BRIR)  was measured with a dummy head using a directional loudspeaker (Genelec 8020c) as a source placed near the dummy head. Fig. \ref{BRIRsetup} shows a photo of the measurement setup inside the empty room, i.e. the only objects in the room were the dummy head and the loudspeaker; the sound interface and the computer were placed outside the room. 
	
\begin{figure}[ht] 
\centering
\includegraphics[width=.45\textwidth]{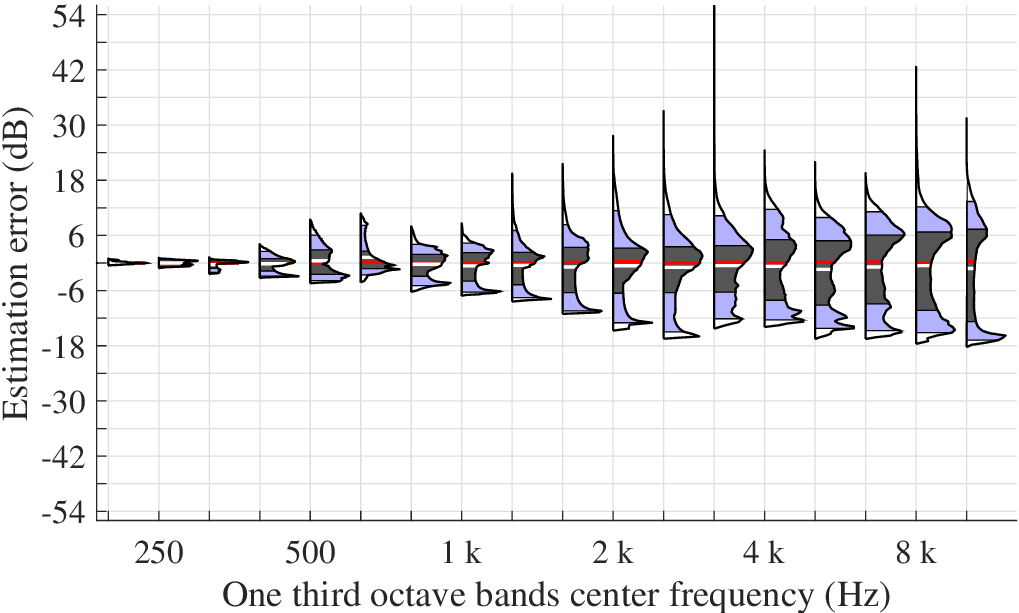}
\caption{Estimated probability density functions of directivity's estimation errors of a Genelec 8020c. The dashed area shows the percentile 25 and 75, the red solid line shows the mean and the white solid line shows the median.\label{fig:genelc_pdf}}
\end{figure}

The probability density functions of directivity estimation errors for the Genelec 8020c are shown in Fig. \ref{fig:genelc_pdf}.
The source data consisting of the radiation patterns in a 1-degree resolution spherical grid are available in \cite{BRAS2019}. The analysis was performed as in \cite{accolti2022}, by averaging the radiation pattern in frequency bands and comparing the resulting value with the radiation pattern for each frequency within each band.

About 50 \% of the errors are within the $\pm 6\ \textrm{dB}$ range for each 1/3 octave band below 4 kHz. This error range increases with frequency likely due to the higher variation of directivity within wider 1/3 octave bands at higher frequencies.

\begin{figure}[ht]
\begin{center}
\includegraphics[width=0.35\textwidth]{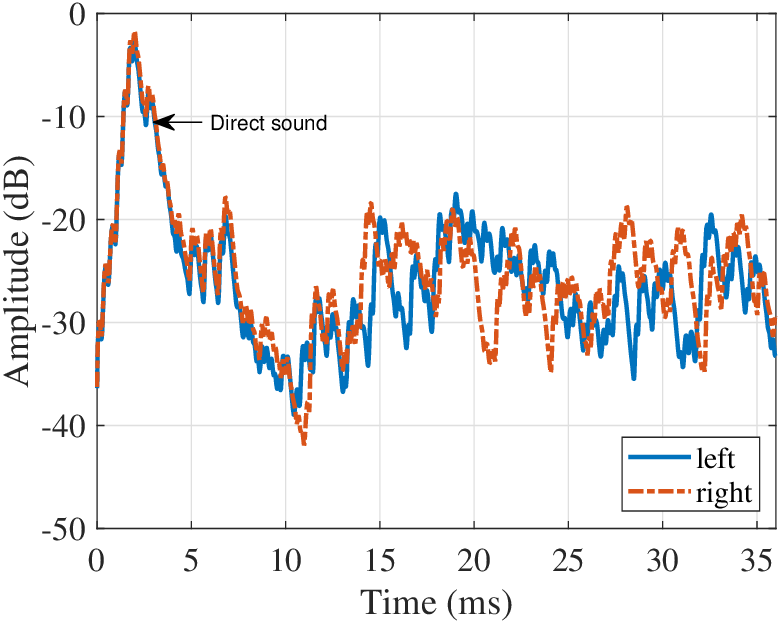} 
\end{center}
\caption{Measured binaural room impulse response. Envelope estimated with  Hilbert transform at mid the frequencies range from 353 Hz to 2828 Hz (i.e. from 500 Hz to 2 kHz Hz octave bands)}%
\label{BRIR_meas}%
\end{figure}

Fig. \ref{BRIR_meas} shows the envelope of the  measured BRIR. It was measured and processed with the ITA-toolbox \cite{dietrichToolboxDAGA2010,ITA-Toolbox_2017}. 
The  envelope was estimated with the Hilbert transform as in \cite{Hidaka_2022} because it provides a good descriptor for analyzing the texture of room impulse responses.

\begin{figure}[h!]
\begin{center}
\includegraphics[width=0.35\textwidth]{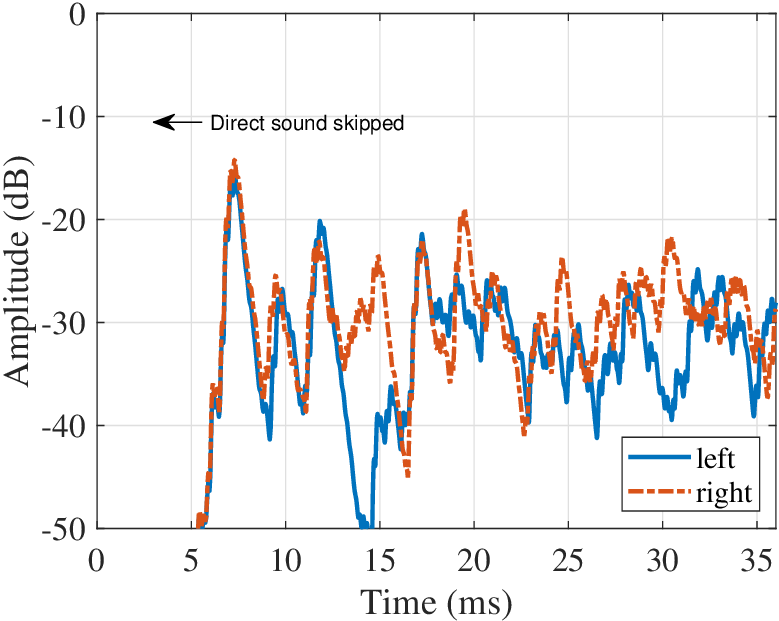}
\end{center}
\caption{Simulated binaural room impulse response. Envelope estimated with Hilbert transform at mid the frequencies range from 353 Hz to 2828 Hz (i.e. from 500 Hz to 2 kHz Hz octave bands)}%
\label{BRIR_simu}%
\end{figure}

The same setup was also simulated using GA modelling using RAVEN \cite{schroder_raven_2011}. 
The measured \text{HRTF} of the same dummy head and the measured directivity of the same loudspeaker (Genelec 8020c) were included as RAVEN input data. 
Then, the resulting BRIR was convolved with the loudspeaker impulse response because the measured directivity of the loudspeakers are normalized to the radiation in the axial direction. 
Fig. \ref{BRIR_simu} shows envelope of the simulated BRIR. 

Comparing the results in Figs. \ref{BRIR_meas} and \ref{BRIR_simu}, the measurement shows a direct sound at $\Delta t \approx 2\ \text{ms}$ that is not present in the simulation because the direct sound is excluded in the hearing oneself case. Then, the first reflection is from the floor for both the measurement and the simulation. Furthermore, the energy of the subsequent early reflections in the simulation are more-or-less similar to that of the measurement. However, these subsequent reflections are not as similar as for the first reflection, which occurs due to the mentioned calibration uncertainties, simulation data inputs, and simulation modeling. The modeling of the sound diffusion is achieved by RAVEN as a  gradual mix of the ray tracing model with the image source model. Hence, the sound energy is similarly distributed in a statistical sense across short time intervals.

The technique for the hearing-oneself case is to keep direct sound as in the measured case (Fig. \ref{BRIR_meas}) and replace the following sound energy with the simulation excluding the direct sound (Fig. \ref{BRIR_simu}). 
This is carried out by placing each musician in a test room with open headphones that allow the direct sound of her or his own instrument to pass through the cups of the headphones and playback the sound processed in real time to simulate the reflections over the headphones. 
The following two subsections explore this technique.

\subsection{Objective evaluation}
\label{subsec:calforguit}

In this subsection, the calibration is estimated and the effect of headphones attenuation on real direct sound are shown for a general case. This case considers the dummy head and the loudspeaker in Fig. \ref{BRIRsetup} placed in a free field environment for emulating a virtual environment for a soloist musician ($M=1$) in hearing oneself configuration ($N=1$). 

\begin{figure}[h!]
	\centering
		\includegraphics[width=0.35\textwidth]{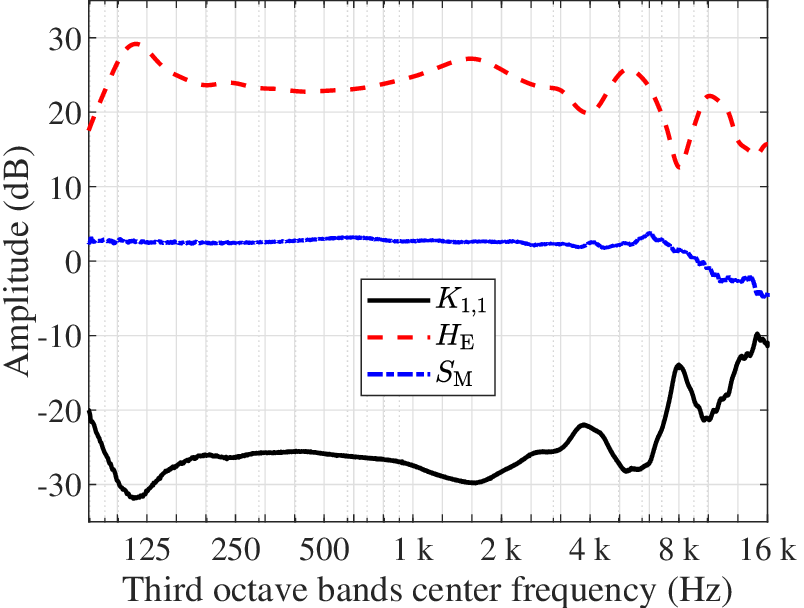}	
	\caption{Calibration for a simulated guitarist in a hearing oneself configuration}
	\label{fig:PC_Self_Guit}
\end{figure}

Fig. \ref{fig:PC_Self_Guit} shows the modulus in decibels of the calibration obtained for this configuration. Besides, it includes $S_{\text{M},m}$ from Fig. \ref{fig:SM} and $H_{\text{E},n}$ from Fig. \ref{fig:SE} as references. 

\begin{table*}[hb] 
\caption{Configuration of the scenarios and relevant acoustic parameters \label{tab:scen_conf}}
\begin{center}
\begin{tabular}{l r r r r r r r}
Scenario           & $\text{R}_1$ & $\text{R}_2$ & 
                                           A &  B & C  &  D  &  M \\
\hline 
Stage height (m)          & 4 to 9  & -- & 9  & 18 &  9 & 18 & 13.5 \\
Stage width (m)           &      12 & 20 &20  & 20 & 12 & 12 & 16 \\
\hline \hline
Reverberation time (s)    & 0.9     & -- & 1.0 & 1.1 & 0.9     & 1.0& 1.0\\
\hline \hline
Early stage support (${ST}_{\text{E}}$; dB) &-12.4& -- & -13.6 & -16.6 & -11.8 & -14.2 & -14.4 \\ 
Late stage support (${ST}_{\text{L}}$; dB) &-18.1& -- & -16.0 & -16.2 & -15.7 & -16.0 & -16.6 \\ 
 \end{tabular}
\end{center}
\end{table*}

The calibration is carried out by eq. \eqref{eq:k} and assuming an error $E_{\text{K},m,n}$ (i.e. replacing $E_{\text{K},m,n}=1$). The microphones used for estimating $H_{\text{E},n}$ and $S_{\text{M},m}$ in eq. \eqref{eq:k} (Sennheiser KE4 and Oktava MK-012, respectively) were calibrated by comparison to a calibrated class 1 measurement microphone in the center of a medium sized room with a reverberation time below $0.2\ \text{s}$~\cite[see room called Virtual laboratory in Sec. 2.3]{pausch2022}). 

\begin{figure}[ht]
	\centering
		\includegraphics[width=0.35\textwidth]{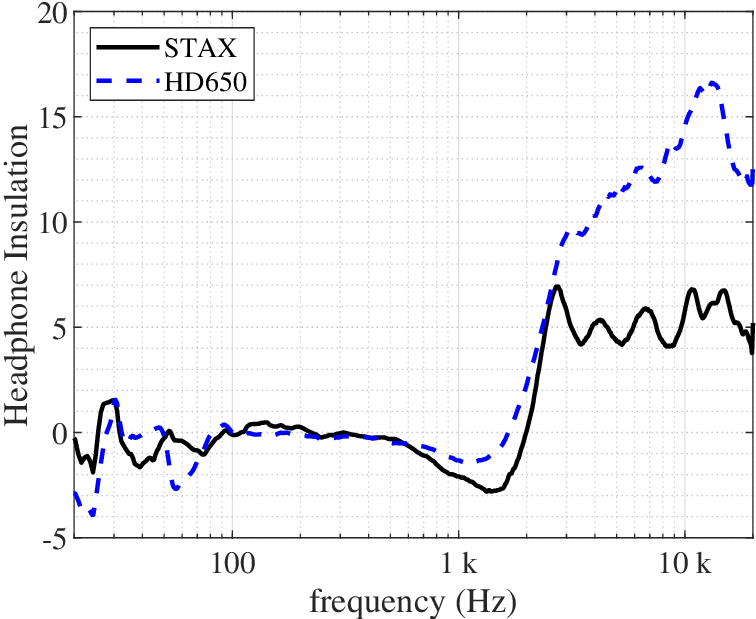}	
	\caption{Attenuation of LAMBDA STAX and Sennheiser HD 650}
	\label{fig:fig_HeadphonesAt}
\end{figure}

The actual direct sound arrives from outside the headphones and just the reflections are simulated through the loudspeakers of the headphones. 
Hence, the attenuation of the headphones is assumed as an error in representing the direct sound for the hearing oneself case (i.e. it does not apply when listener $n$ hears musician $m\neq n$). Fig. \ref{fig:fig_HeadphonesAt} reports the sound attenuation measured for LAMBDA STAX and Sennheiser HD 650 with a dummy-head and a loudspeaker in twelve equidistantly distributed positions in a sphere of 2.5 m radius. LAMBDA STAX headphones were used for the following proof of concepts as they involve lesser attenuation of the direct sound.

\subsection{Subjective evaluation}

In order to provide subjective support, an experiment with six actual guitar duets was conducted. Each duet interacted over several simulated scenarios. Per scenario, they also evaluated the overall quality, and the similarity to playing in a real room. 

The twelve guitarists' age was 30.4 years in average (4.8 years standard deviation) with self-reported normal hearing. Although only one of them is a professional musician, every guitarist reported some experience with guitar ensemble either during the past up to ten years or during the current year. Musicians received a monetary compensation.

\begin{figure}[ht!]
    \centering
    \subfloat[Floor plan scheme\label{fig:plan}]{\includegraphics[width=0.38\textwidth]{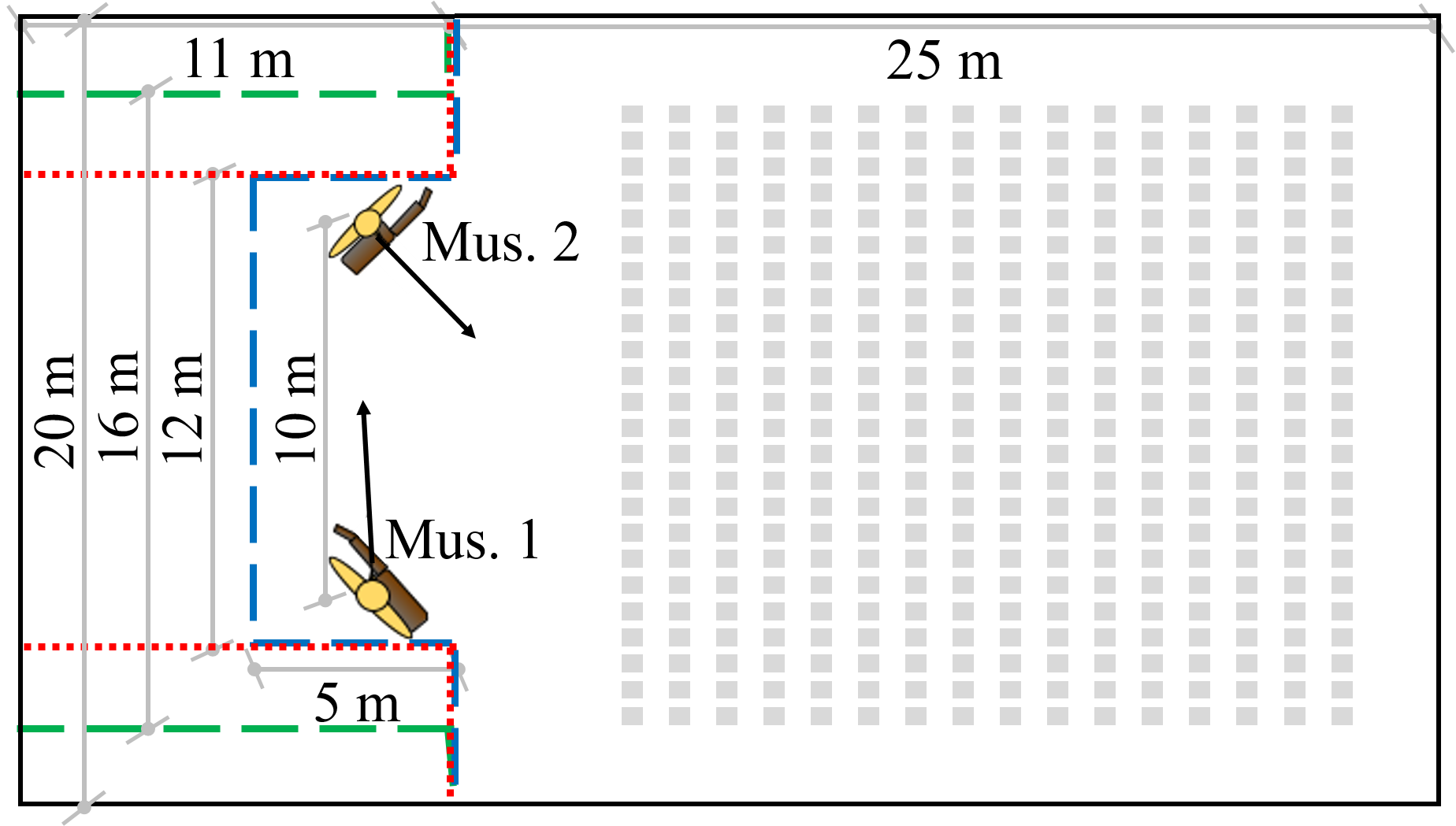}}
    \hfill
  \subfloat[Longitudinal section scheme\label{fig:section}]{%
        \includegraphics[width=0.38\textwidth]{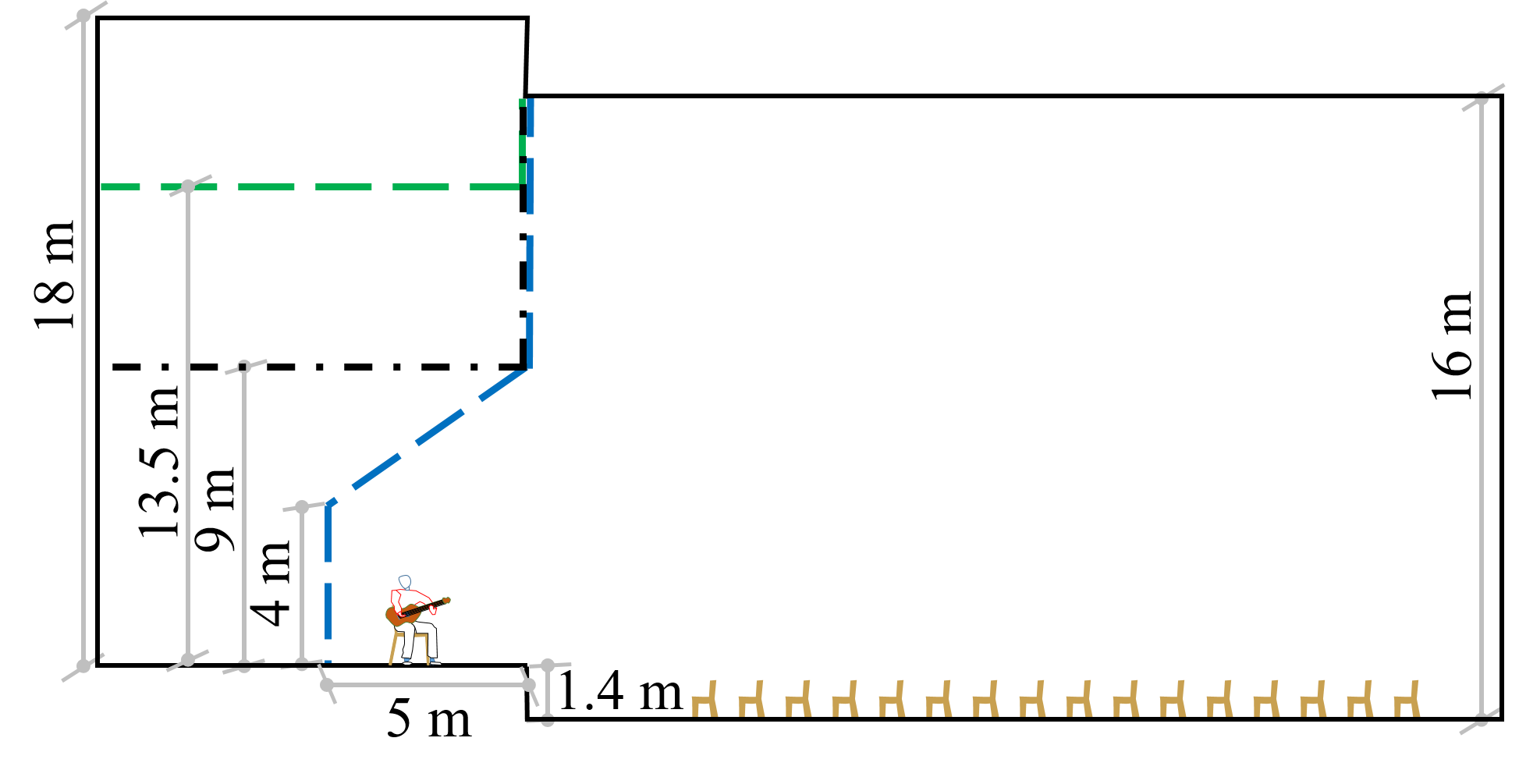}}
    \\
    
    \caption{Scheme of the virtual scenario. Solid line: bigger room with high level of height and width. Red dotted line: low level width. Black dash-dotted line: low level height. Blue long dashed line: tight-shell}
    \label{fig:rooms}
\end{figure}

Each duet performed in eight different scenarios, whose main characteristics are shown in Fig.~\ref{fig:rooms}. The eight scenarios have the same audience plan dimensions ($25\ \text{m} \times 20\ \text{m}$) and height (16\ \text{m}). The  width and height of stages are shown in Tab. \ref{tab:scen_conf}. 

The first two scenarios ($\text{R}_1$ and $\text{R}_2$) were provided for the musicians to get familiar with the experiment. 
The first scenario $\text{R}_1$ had a tight stage shell (see blue long dashed line in Fig.~\ref{fig:rooms}) which provides acoustic support (with walls 1 m apart from musicians and a splayed ceiling) and the second one $\text{R}_2$ is  just the stage and the audience surfaces as in an outdoors situation (i.e. without any walls or ceiling). 
This was followed by the six experimental scenarios (A, B, C, D, and two times M) in random order. The experimental scenarios are shoebox-shaped stages whose dimensions were set by the four possible combinations (A, B, C, and D) of a $2^k$ statistical design varying two factors $k$ at two levels plus a central point (M) with a repetition. The statistical design involves two levels of width (9 m and 18 m, see red dotted line and solid line, respectively in Fig.~\ref{fig:rooms}), two levels of height (12 m and 20~m, see black dash dotted line and solid line, respectively in Fig.~\ref{fig:rooms}), and one level of depth (11 m).

 Tab. \ref{tab:scen_conf} also shows acoustical descriptors of the virtual scenarios. Due to identical audience simulation (i.e., geometry and materials are the same), the reverberation time is similar for all the scenarios except for the free-field case $\text{R}_\text{2}$.  
Early stage support ${ST}_{\text{E}}$ and late stage support ${ST}_{\text{L}}$ are reported in Tab. \ref{tab:scen_conf}~\cite{iso3382}. 
They are the average values for octave bands from 250 Hz to 2 kHz and for three source positions, each with three receiver positions. 

The scenarios' dimensions were chosen in order to have some diversity in stage acoustic quality as reported in previous work on stage acoustics of real rooms~\cite{panton_chamber_2019}. Besides, $ST_{\text{E}}$ range of the scenarios is $[-11.8 , -16.6]$ with about 5~dB difference between scenarios B and C. $ST_{\text{E}}$ of scenarios A and D are similar  with about 0.6~dB difference.

The distance between guitarists was set at a constant virtual distance of 10 m, i.e., $d_{\text{S-E},{1,2}}=d_{\text{S-E},{2,1}}=10\ \text{m}$ between musicians. This was primarily to ensure that the differences between experimental conditions were easier to perceive. 
Smaller distances between musician increase the ratio of direct  to reflected field, in turn making reflected field harder to be perceived. 
The latter was also confirmed during pre-experiment explorations where for distances more typical in real ensembles (e.g., around 2 m), it was hard to notice differences in experimental variations. This does not imply that perceiving differences in stage acoustics is hard in general, but just that they were relatively harder to perceive with shorter inter-musician distances for some of the current set of experimental conditions. 

For each duet, the two musicians were located in separate rooms connected only aurally by their instruments (i.e., with microphones and headphones implementing eq. \eqref{eq:soj}).
The microphones were Oktava MK-012 and the headphones were STAX LAMBDA (Figs. \ref{fig:SM}, \ref{fig:SE}, and \ref{fig:fig_HeadphonesAt} show the microphone sensitivity, the HpTF, and the headphones cups' attenuation, respectively). 
Besides, the source\hyp{}microphone distance was $\hat{d}_{\text{M-S},m}=0.9\ \text{m}$ for both guitarists $m\in\{1,2\}$. 

Musicians were informed that they would play on a virtual stage, positioned 3 m from the edge of the stage and 10 m apart from each other. Since guitar duets usually play with shorter inter-musician distances (roughly 2 m apart typically), they were instructed to imagine that the 10 m distance was due to them being part of a bigger ensemble.  
They were also informed on their virtual orientations, which were quite different for each musician (Fig.~\ref{fig:plan}).
Looking from the stage to the audience, the head of the musician at the right (Mus. 1) is turned left to the fretboard and the head of the musician at the left (Mus. 2) is directed to the audience in front and between the fretboard and the position of the other musician\footnote{This setup for the orientation of the receivers was decided based on recorded videos of several professional duets. 
All participants were right-handed players; hence no other configuration was used.}.

Participant duets were asked to play two short  (about 60 s long) musical pieces for guitar ensemble (one of them allegro and the other one adagietto or andante) in the simulated scenarios. 
Although the participants communicated between the scenarios using their voices, they were asked to base their judgments just on guitar signals, acoustics, and ensemble playing, and not on visuals.  
To that end, they also tested the scenarios with their guitars playing single tones, chords, arpeggios, or improvising at the beginning and the end of each scenario.

After playing in each scenario, musicians answered a questionnaire that included items, each on a linear scale, about aspects related to the acoustical quality. While most of the questionnaire items will be analyzed in future work, an item asking about the 'overall quality' ('bad acoustics' -- 'good acoustics' at the extremes of the scale) per scenario is analyzed here. Since distributional assumptions of parameteric tests were not met, non-parameteric tests are used throughout.

\begin{figure}[h!]
    \centering
    \includegraphics[width=.35\textwidth]{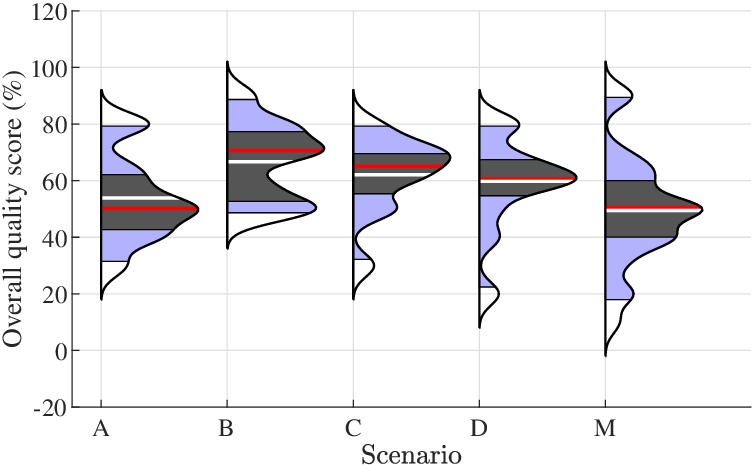}
    \caption{Probability density function of the Overall quality for  each scenario (4 \% Gaussian kernel)}
    \label{fig:OQ_s}
\end{figure}

Wilcoxon signed-rank test showed that the participants' ratings for the two instances of condition M were not significantly different ($Z$ = -0.83, $p$ = .41). 
This indicates that participants provided similar ratings to repeated instance of the same condition, which is at least an encouraging sign for the subjective evaluation of the system.
Henceforth, given their similar values, the ratings for repetitions of condition M were averaged. 
Following this, Friedman's test indicated that the overall effect of experimental condition over ratings was statistically significant $(\chi^2 (4)= 11.14, p = .025)$. Fig.~\ref{fig:OQ_s} shows the PDFs for each condition. Finally, post-hoc tests (Wilcoxon test with $p$--value adjustment using the FDR method) showed that none of the conditions were statistically significant from each other (all $p > $  .5). 
To summarize, while there is an overall effect of experimental condition, pairwise or other comparisons (e.g., between different groups of heights and widths) did not reach statistical significance. There can be many reasons for the latter, including small sample sizes (limiting the statistical power) which is a clear limitation of the current experiment, and the overall significance being driven by specific group differences that are not meaningful comparisons based on experimental design, etc. However, from the perspective of the system evaluation, since there are meaningful trends among conditions, the results seem promising for future testing with a larger sample size. 

\begin{figure}[ht]
    \centering
    \includegraphics[width=.40\textwidth]{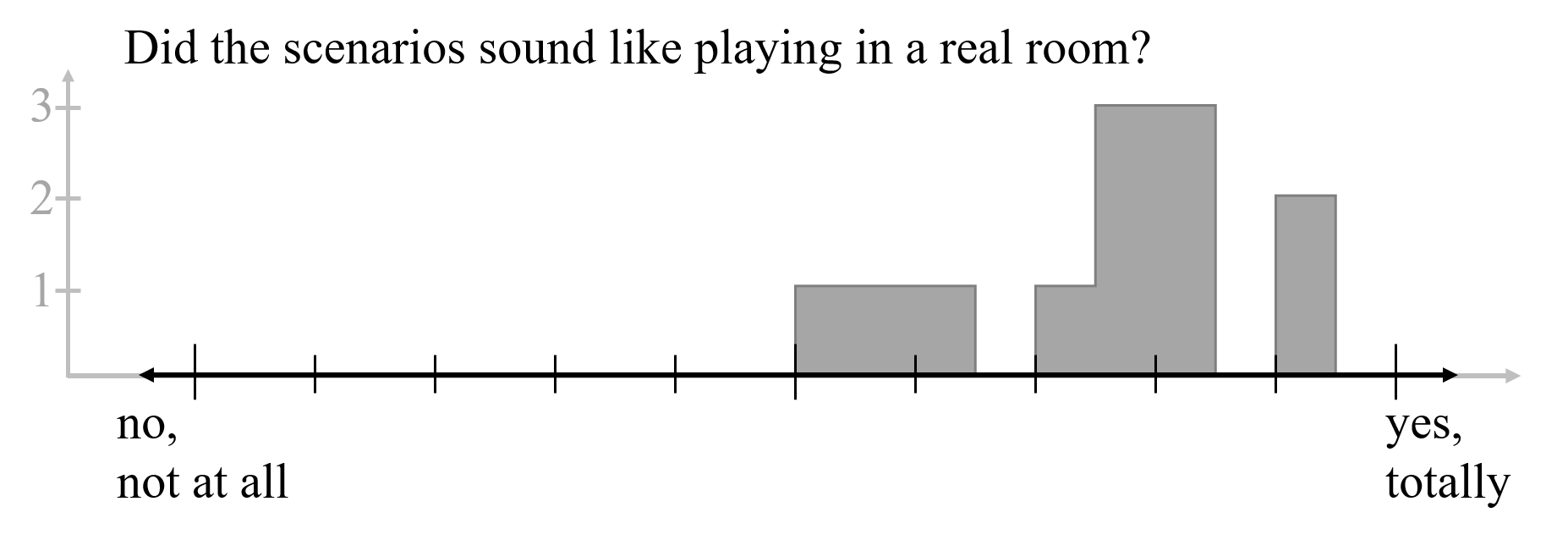}
    \caption{Subjective judgement of the 12 participants. Black: presented scale. Grey: histogram of collected responses}
    \label{fig:result}
\end{figure}

Finally, after playing in all the proposed scenarios, a linear scale was presented to collect responses on the similarity of playing in the proposed scenarios to playing in a real room. 
An histogram of the collected responses is shown in Fig.~\ref{fig:result}. Let the extreme values of the scale be 0 and 100, respectively; then, the mean is $73$, the mode is $80$, and the standard deviation is $13$. 

\section{Discussion}
\label{sec:discussion}

The findings indicate that the presented real-time system for auralizing musical performance of two or more musicians interacting in an ensemble is feasible. Hence, the presented system substantially extends the possibilities of multi-musician auralizations compared to existing systems for one musician interacting in real-time with recorded signals \cite{p-g_equal_2011, y_c_m_2012}. A key advantage of the presented system is the increased interactivity: musicians can modify their performances due to the real-time ensemble. This represents a more dynamic and closer-to-reality playing condition compared to interacting with recordings.

The critical latency issue for the hearing oneself configuration (about less than 3 ms) is addressed in the presented system by avoiding the digital simulation of direct sound. This technique, which has previously been used in non-interactive systems (e.g., \cite{p-g_equal_2011,y_c_m_2012}),  allows for a larger latency of about 6~ms considering a 2~m distance between musicians.

On the one hand, some latency may not interfere with tempo of the musical performance, although it may change the perceived room acoustics. Hence, proper latency compensation is an important consideration in systems intended for stage acoustics studies. 
On the other hand, the drawback with not simulating the direct sound is that there will be some direct sound attenuation due to the headphones. 
To minimize this effect, headphones with relatively smaller attenuation (about 6 dB at frequencies higher than 2 kHz; Fig. \ref{fig:fig_HeadphonesAt}), compared to an alternative, were used in the subjective evaluation of this paper. Further, in the case of a guitarist in a hearing oneself configuration, the quickest reflection comes from the guitar\hyp{}floor\hyp{}ear path, which was accurately simulated with a suitable off-the-shelf hardware interface. It is possible that with more customized hardware, the latency issues could be curtailed further in future implementations.

Besides the latency issue, the implications of the calibration filter were analyzed. 
To that end, a comprehensive theoretical derivation of the components of a calibration filter for the system was presented.
It was determined that the three main sources of error are: the directivity of sources, the distances from source to the microphone to the head,  and the transfer function of headphones.
Although not considered in this paper, head movement of the guitarists may have an effect on these three sources of error (i.e. the headphones transfer function may be slightly modified if the headphones move during performance).

The uncertainty due to band averaging the source directivity is comprehensively analyzed elsewhere \cite{accolti2022}. However, it is important to note that, as far as the authors know, no study has investigated the directivity differences among units of the same instrument. Furthermore, each musician played a different guitar in the experiments because musicians were encouraged to play with their own instruments for comfort. This is a clear weakness of this study and other similar studies. Hence, investigating this uncertainty further is an open research topic.

The effects of the single number estimation of source\hyp{}microphone and source\hyp{}head distances derived in this paper (Fig.~\ref{fig:error} also apply for listeners in the audience of performance spaces, conference halls, or other rooms. 
This effect grows with relative errors in distances. 
Therefore, larger real source\hyp{}microphone and virtual source\hyp{}head distances are less sensitive to this effect.

The objective proof of concepts showed encouraging results, hence a subjective experiment was carried out. 
The subjective proof showed that playing in 
the simulated scenarios was judged as having a score above the middle of a scale measuring similarity to playing in a real room (Fig.~\ref{fig:result}) by all the participants. 
Furthermore, for the more experienced musicians, this rating was closer to the maximum score.

The other weaknesses of this setup are related to the lack of visual feedback and the lack of response to source and head movements (especially for head movements of the guitarist whose left hand is closer to the side wall). Improving both these weaknesses is challenging due to latency issues as discussed in the following two paragraphs.

On the one hand, preliminary experiments carried out previous to the proof of concepts with experienced guitarist participants showed that static visual feedback did not improve the ratings of similarities to playing in a real room (Fig.~\ref{fig:result}). On the other hand, latency compensation of real-time video feedback is challenging. A latency mismatch between video and audio could introduce further uncontrolled variables, which was avoided here by excluding visual feedback altogether. Visual feedback can be improved by positioning musicians in the same room or in separated rooms with glass windows, hence visually connecting them like in recording studios. However a recording-studio-like setup could also interfere with aural cues of distance between musicians. Nevertheless, improving audio-visual integration represents future work.

Although not studied in this article, placing two or more musicians in the same room may be useful to reduce the distance between them to be below 2~m. Of course, in this case the direct sound would be skipped in the room impulse responses for hearing others (i.e. $m\neq n$). This, however, could not be achieved in the current implementation with musicians playing in separated rooms. Moreover, placing musicians in the same room reduces the flexibility in varying the distances between them across experimental scenarios, and/or in setting virtual distances beyond the dimensions of the actual laboratory rooms; both of these are possible within the current system.

Tracking head movements represents the current state-of-the-art technique for auralization of prerecorded material. However, when the audio signals are generated just 6 ms before, the tracking of movements becomes a challenging task.

Despite these limitations, the system was able to provide a realistic and engaging listening and playing experience. The results of this study suggest that real-time auralization of musical performance is a promising technology with the potential to provide experimental platform for stage acoustics investigations as well as musical performance  training. 

\section{Conclusions}
\label{sec:conclusions}

An interactive and real-time system for stage acoustics experimentation is proposed and validated. The system enables interactive listening, including hearing oneself as well as hearing others, while playing simultaneously on a virtual stage. The latency issues are circumvented by time compensation of the simulated impulse responses. The main sources of uncertainty are discussed based on state-of-the-art practices and databases. Future studies are recommended to address knowledge gaps regarding source directivities including their interaction with typical movements by musicians during playing. The results from a pilot study with guitar duets show promise in terms of feasibility of the system for duets. Moreover, the current system can be scaled to accommodate a larger number of musicians and musical instruments. This can be useful in enabling virtual stage acoustics studies with a large number of musicians concurrently, which currently represents a challenge. 

\section*{Acknowledgment}
{The work of E. A. in Germany was supported by the Alexander von Humboldt Foundation}.
{M.Y. was supported by a DFG (German Research Foundation) grant - Project number 503914237}.

\bibliographystyle{elsarticle-num} 

\bibliography{ST-DIF}

\end{document}